\documentclass[twocolumn,aps,
prb,showpacs,letterpaper,superscriptaddress]{revtex4-1}

\usepackage[dvips]{graphicx,color}

\newcommand{\Eqref}[1]{Eq.~(\ref{#1})}
\newcommand{\Secref}[1]{Sec.~\ref{#1}}
\newcommand{\Figref}[1]{Fig.~\ref{#1}}

\newcommand{\revision}[1]{{#1}}

\begin{document}

\title{
Simulation of inelastic electron tunneling spectroscopy of
single molecules \\with functionalized tips 
}

\date{\today}

\author{A. Garcia-Lekue}
\affiliation{Donostia International Physics Center (DIPC), Paseo Manuel de
Lardizabal 4,
E-20018 San Sebastian, Spain}
\author{D. Sanchez-Portal}
\affiliation{Donostia International Physics Center (DIPC), Paseo Manuel de
Lardizabal 4,
E-20018 San Sebastian, Spain}
\affiliation{Centro de Fisica de Materiales CFM-MPC, 
Centro Mixto CSIC-UPV/EHU, Paseo Manuel de
Lardizabal 5,
E-20018 San Sebastian, Spain }
\author{A. Arnau}
\affiliation{Donostia International Physics Center (DIPC), Paseo Manuel de
Lardizabal 4,
E-20018 San Sebastian, Spain}
\affiliation{Centro de Fisica de Materiales CFM-MPC, 
Centro Mixto CSIC-UPV/EHU, Paseo Manuel de
Lardizabal 5,
E-20018 San Sebastian, Spain }
\affiliation{Depto. de Fisica de Materiales UPV/EHU, Facultad de Quimica, Apdo.
1072, San Sebastian, Spain}
\author{T. Frederiksen}
\affiliation{Donostia International Physics Center (DIPC), Paseo Manuel de
Lardizabal 4,
E-20018 San Sebastian, Spain}

\begin{abstract}

\vspace*{1mm}

The role of the tip in inelastic electron tunneling spectroscopy (IETS) performed with
scanning tunneling microscopes (STM) is theoretically addressed via
first-principles simulations of vibrational spectra of single 
carbon monoxide (CO) molecules adsorbed on Cu(111). We show how chemically 
functionalized STM tips modify the IETS intensity corresponding to
adsorbate modes on the sample side.
The underlying propensity rules are explained using symmetry considerations for both the vibrational modes 
and the molecular orbitals of the tip and sample.
This suggests that single-molecule IETS can be optimized 
by selecting the appropriate
tip orbital symmetry.

\end{abstract}

\maketitle

\section{Introduction}

Vibrational spectroscopy with the scanning tunneling microscope (STM) is
a powerful technique to study molecules 
adsorbed on surfaces.\cite{stipe_science98}
The method combines the high spatial resolution of the STM with
the chemical sensitivity of 
inelastic electron tunneling spectroscopy (IETS) as first reported
for molecules incorporated in metal-oxide-metal 
tunneling junctions.\cite{JaLa.66.MolecularVibrationSpectra} 
At low temperatures, molecular vibrations are almost completely frozen 
but can be excited by tunneling electrons with sufficient energy. By
recording the changes in tunneling conductance 
as a function of applied bias voltage
the onsets of distinct phonon emission 
processes can be detected. In this way spectra recorded on single molecules are 
vibrational fingerprints that reveal information about chemical species 
and local environment. Inelastic tunneling electrons from the STM can also
be used to induce and control molecular motion\cite{KoKiKa.02.Lateralhoppingof,
HeLuGu.2002.MoleculeCascades,PaLoSo.03.Selectivityinvibrationally,
LoRuTa.05.Single-moleculemanipulationand}
and single-molecule chemistry.\cite{Ho.02.Single-moleculechemistry}

Several theoretical methods have been developed to describe 
IETS-STM,\cite{appelbaum_PRB69,caroli_JPC72,
PEBA.87.INELASTICELECTRON-TUNNELINGFROM,bonca_PRL95,emberly_PRB00,
GaRaNi.04.Inelasticelectrontunneling} 
including approaches based on \emph{ab initio} 
calculations for many-electron systems.\cite{mingo_PRL00,lorente_PRL00,
FrBrLo.04.InelasticScatteringand,AlSaAr.10.Mixed-ValencySignaturein}
Good quantitative agreement between simulated and experimental 
data have been reported for many systems,\cite{vitali_nano10,
OkPaUe.10.InelasticTunnelingSpectroscopy,ArFrRu.2010.Characterizationofsingle-molecule,lorente_PRL01} 
but the spectra are difficult to rationalize
due to the complexity of the underlying symmetries of electronic
structure and vibrational modes.
Therefore, in order to identify the active vibrational 
modes in IETS and to understand the corresponding magnitude and sign 
of the conductance changes various
theories of inelastic \emph{propensity rules} 
have been developed.\cite{lorente_PRL01,TrRa.06.Moleculartransportjunctions,
GaSoPe.07.priorimethodpropensity,paulsson_PRL08} 
The many-body extension of the Tersoff-Hamann approach developed by 
Lorente and Persson \cite{lorente_PRL00} permits the derivation of 
selection rules based on the symmetry of the vibrational modes and the 
molecular orbitals close to the Fermi level.\cite{lorente_PRL01} 
However, the morphology and chemical composition of the STM tip is often largely unknown
in the experiments and, therefore, tip effects can be important for the interpretation 
of experimental data.\cite{BaMeRi.99.evolutionofCO,TePeAr.07.Includingprobetip,Calleja10} 
For example, chemically functionalized tips were shown to reveal vibrational modes 
which are not observed with a bare metallic tip.\cite{hahn_PRL01}
Consequently, understanding and disentangling the role of the STM tip in 
IETS measurement are of relevance. 

In this paper we investigate theoretically the interplay between symmetries 
imposed by the electronic structure of the STM tip and the IETS spectra
originating from a vibrating molecule adsorbed on the metal sample.
Specifically, we consider the prototype system of a CO molecule adsorbed on 
Cu(111) in combination with bare and chemically functionalized STM tips.
This choice is motivated by the fact that this system has been the subject
of research with a variety of surface analysis techniques for decades 
and that the observed IETS with metallic tips is 
now well understood.\cite{LaHo.1999.Single-moleculevibrationalspectroscopy,
Pe.04.Theoryofelastic,vitali_nano10}
We show that chemically functionalized STM tips
can increase the resolving power of IETS and suggest
the importance of the symmetries imposed
by the electronic structure of the tip probe.

The outline of the paper is as follows.
In \Secref{theory} we describe the 
methodology used in the present study to simulate IETS. 
In \Secref{results} we report our calculations for 
several specific junctions. In a first set of 
calculations (\Secref{symmetric})
we examine aligned junctions in which 
the sample CO and the molecule on the tip side
share a common rotation axis. 
These results are analyzed in terms of the symmetry and spatial 
distribution of the most transmitting eigenchannels
of the junction in order to establish the inelastic propensity rules. 
Then, in \Secref{asymmetric}, we consider junctions
in which the tip has been laterally displaced. This relaxes
the symmetry constraints and, therefore, increases the 
number of allowed transitions contributing to the IETS signal.
We find that the symmetry of the scattering states
can still be used to understand the IETS signal.
However, rather
than the global symmetry of these, it is 
the {\it local} symmetry of the 
tip scattering states in the region where the 
deformation potential is significant that determines
whether or not they are involved in the inelastic
scattering. 
Finally, in \Secref{sec:conclusions} we summarize
our study and formulate some general predictions
for the influence of the tip symmetry in IETS.

\begin{figure}[t]
\begin{center}
\includegraphics[width=\columnwidth,angle=0]{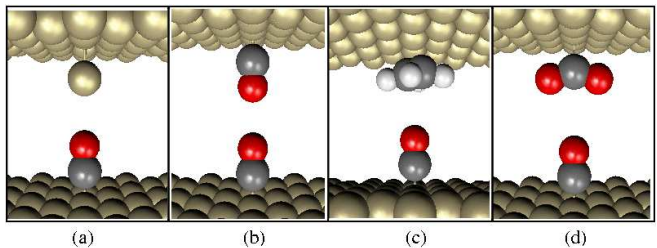}
\caption{\label{fig1}
STM setup: CO molecule on Cu(111) with (a) a Cu adatom tip, or
with a tip functionalized with (b) CO, (c) C$_2$H$_4$,
or (d) CO$_2$. Throughout this paper the lower
surface with the CO molecule is considered the STM sample
side and the upper surface the tip side.}
\end{center}
\end{figure}

\section{Theory}
\label{theory}

We use density functional theory (DFT) combined 
with non-equilibrium Green's function (NEGF) methods
to perform IETS-STM simulations.
The generic systems considered for the simulations are
shown  in \Figref{fig1}. We use a supercell description
of the scattering region with a $3\times3$ representation of a slab containing 6 
Cu(111) layers.
To represent the STM sample side, we place a single CO molecule adsorbed on
a top-site of one of the Cu(111) surfaces. On the opposite surface,
we place a single Cu adatom [\Figref{fig1}(a)], 
 a carbon monoxide [CO, \Figref{fig1}(b)], an ethylene
 [C$_2$H$_4$, \Figref{fig1}(c)]  or a carbon dioxide [CO$_2$, \Figref{fig1}(d)] molecule to model
the STM tips.

The electronic structure and geometries are calculated with the 
\textsc{Siesta} code.\cite{siesta} 
Atomic coordinates of the CO molecule, tip, and surface
copper atoms
were relaxed until forces were smaller than $0.02$\,eV/\AA.
The basis set consisted of single-zeta plus polarization orbitals for Cu atoms
and split-valence double-zeta plus polarization orbitals
for C, O, and H atoms. 
We used the generalized gradient approximation for 
exchange-correlation,\cite{PeBuEr.96.Generalizedgradientapproximation}
a cut-off of 200 Ry for the real-space grid integrations, and
the $\Gamma$-point approximation for sampling of the
three-dimensional Brillouin zone.

The transport properties are simulated with the \textsc{TranSiesta} setup,\cite{brandbyge_PRB02}
in which the scattering region described above is coupled to two semi-infinite 
electrodes on the sample ($S$) and tip ($T$) sides, respectively.
Since we are only interested in the low-bias regime it suffices to 
calculate the electronic structure in equilibrium. \revision{To keep the
present analysis of the inelastic scattering rates transparent we
also restrict the transport calculations to the $\Gamma$-point
of the two-dimensional Brillouin zone.} Eigenchannel
visualizations are calculated according to the method
described in Ref.~\onlinecite{PaBr.07.Transmissioneigenchannelsfrom}.

To simulate the inelastic corrections to the tunneling current we employ the
\textsc{Inelastica} package.\cite{Inelastica,frederiksen_PRB07}
The vibrational modes $\lambda$ and energies 
$\hbar\omega_\lambda$ are obtained by
diagonalizing the dynamical matrix extracted from finite differences
for atoms in a predefined vibrational region. 
Here, we consider two cases: (i) a realistic situation where 
the sample CO molecule, the tip structure, and the Cu atoms
in the topmost layers of Cu(111) surfaces are considered as vibrationally active
components of the system, or (ii) an imaginative but illustrative situation
where only the sample side (CO and topmost Cu layer) vibrates and the tip side atoms are frozen.
To each vibrational mode we determine the corresponding electron-phonon 
($e$-ph) coupling matrix $\mathbf{M}^\lambda$ via finite
differences of self-consistent Kohn-Sham Hamiltonians.\cite{frederiksen_PRB07}

The inelastic scattering is addressed using the NEGF
formalism within the lowest-order-expansion
(LOE) scheme,\cite{paulsson_PRB05,viljas_PRB05,HaNoBe.10.Currentnoisein}
which relies on weak $e$-ph interactions and the electronic
structure at the Fermi level only. Ignoring quasi-elastic 
correction terms\cite{viljas_PRB05,HaNoBe.10.Currentnoisein}
we write the inelastic current as:
\begin{eqnarray}\label{eq1}
\displaystyle I^\mathrm{LOE}   & = &   G_0\,V\tau \nonumber  \\
\displaystyle & & \displaystyle + \sum_{\lambda} I_{\lambda}^\mathrm{sym} (V,kT,\left<n_\lambda\right>)
 \text{Tr}\big[{\bf G^{\dagger}\Gamma_\mathit{S}\,G}
  \big\{{\bf M^\lambda\,A_\mathit{T}\,M ^\lambda}\nonumber \\
 &&  + {i\over 2}\big({\bf \Gamma_\mathit{T} G^\dagger\,M^\lambda\,A\,M ^\lambda}
- \text{H.c.}\big) \big\}
\big] \nonumber \\
  && + \displaystyle
\sum_{\lambda} I_{\lambda}^\mathrm{asym} (V,kT)
 \text{Tr}\big[{\bf G^{\dagger}\Gamma_\mathit{S}\,G}  \nonumber \\
 & & \times \big\{{\bf \Gamma_\mathit{T}\,G^\dagger\,M^\lambda
(A_\mathit{T}-A_\mathit{S})M^\lambda  +\text{H.c.} }\big\}
\big].
\end{eqnarray}
The first term in \Eqref{eq1} is the elastic Landauer result with
transmission $\tau$ times the conductance
quantum $G_0 = 2e^2/h$. The second and third terms
contain the inelastic corrections to the current from each
vibrational mode at the corresponding threshold voltages.
The functions $I_{\lambda}^\mathrm{sym} \left(V,kT,\left<n_\lambda\right> \right)
$ and  $I_{\lambda}^\mathrm{asym} \left(V,kT\right)$ give rise to
symmetric and asymmetric features in the differential conductance $dI/dV$
and depend on bias voltage $V$, fundamental temperature $kT$,
and thermal expectation value of the mode occupation $\left<n_\lambda\right>$.
The retarded Green's function ${\bf G}$, the spectral function
${\bf A} = i({\bf G-G^\dagger})$, as well as the electrode couplings
${\bf \Gamma}_{S,T}$ are all evaluated at the Fermi energy. The quantities
${\bf A_\mathit{S,T} = G\,\Gamma_\mathit{S,T}\,G^\dagger}$ are 
partial spectral functions defined such that
${\bf A = A_\mathit{S} + A_\mathit{T}}$.
The sum in \Eqref{eq1} runs over all modes $\lambda$ in the
vibrational region.

From the inelastic current \revision{$I(V)$} the IETS, defined as
\begin{equation}\label{eq2}
\text{IETS} \equiv \frac{d^2I\,/dV^2}{dI/dV},
\end{equation}
can be calculated.
\revision{The normalization by $dI/dV$ is convenient to eliminate 
the exponential dependence on the tip-substrate distance of the 
tunneling current. In this way the normalized IETS is a measure of
the inelastic cross sections of the various modes.}
Examination of the ``asymmetric" term in \Eqref{eq1} shows that
it is zero for symmetric systems.\cite{paulsson_PRB05} Experimental IETS curves do
often contain asymmetric signals, although the size of these
signals is usually small.
Such inelastic features
can be removed by
averaging the IETS signals for positive and negative bias.
This symmetrization procedure has been applied
to all the IETS spectra reported in \Secref{results}.
Furthermore, our spectra are broadened according to 
a lock-in signal $V_\mathrm{rms}=1.8$ mV and a temperature 
of $T=6$ K.\cite{frederiksen_PRB07}
\revision{We note that while the bias inversion symmetry 
is only approximate for asymmetric junctions, this symmetry is unrelated to 
the orbital and vibrational symmetries discussed below for the IETS 
propensity rules.}

Our results are 
useful to understand  the effect of the symmetry
on the changes in the strength of the IETS  signals.
With the aim of elucidating the origin of these changes,
we have thoroughly investigated the
role of the tip electronic structure on the inelastic scattering
associated with the vibrational modes of the CO molecule on the sample side.
Nevertheless, caution should be exercised in the quantitative comparison of
our simulated IETS spectra with experiments, since various
approximations are involved in our calculations, such as
$\Gamma$-point sampling and LOE approach.
Although for most
cases our approximations are reasonable, in some
situations they might affect position and amplitude of the IETS peaks.

\section{Results}\label{results}

In our calculations we have considered the
different  junctions shown in \Figref{fig1}.
The sample is always a flat Cu(111) surface with an adsorbed
CO molecule.
The molecule is assumed to adsorb vertically on a top-site of Cu(111),
as experimentally observed. Although such assumption  overcomes some of the 
limitations of many DFT
functionals which incorrectly give the fcc-hollow as the most stable
configuration,\cite{lopez_SS01,alcantara_PRB09}
it introduces inaccuracies in the prediction of the low-energy
frustrated translation (FT)
mode.\cite{vitali_nano10}
Therefore, in the following
we will focus on the 
frustrated rotation (FR) 
and other higher-lying 
vibrational modes,  well-defined in 
our DFT-NEGF calculations.\cite{vitali_nano10}

\begin{figure*}[t]
\centerline{
\begin{tabular}{cc}
\includegraphics[height=6.5cm]
{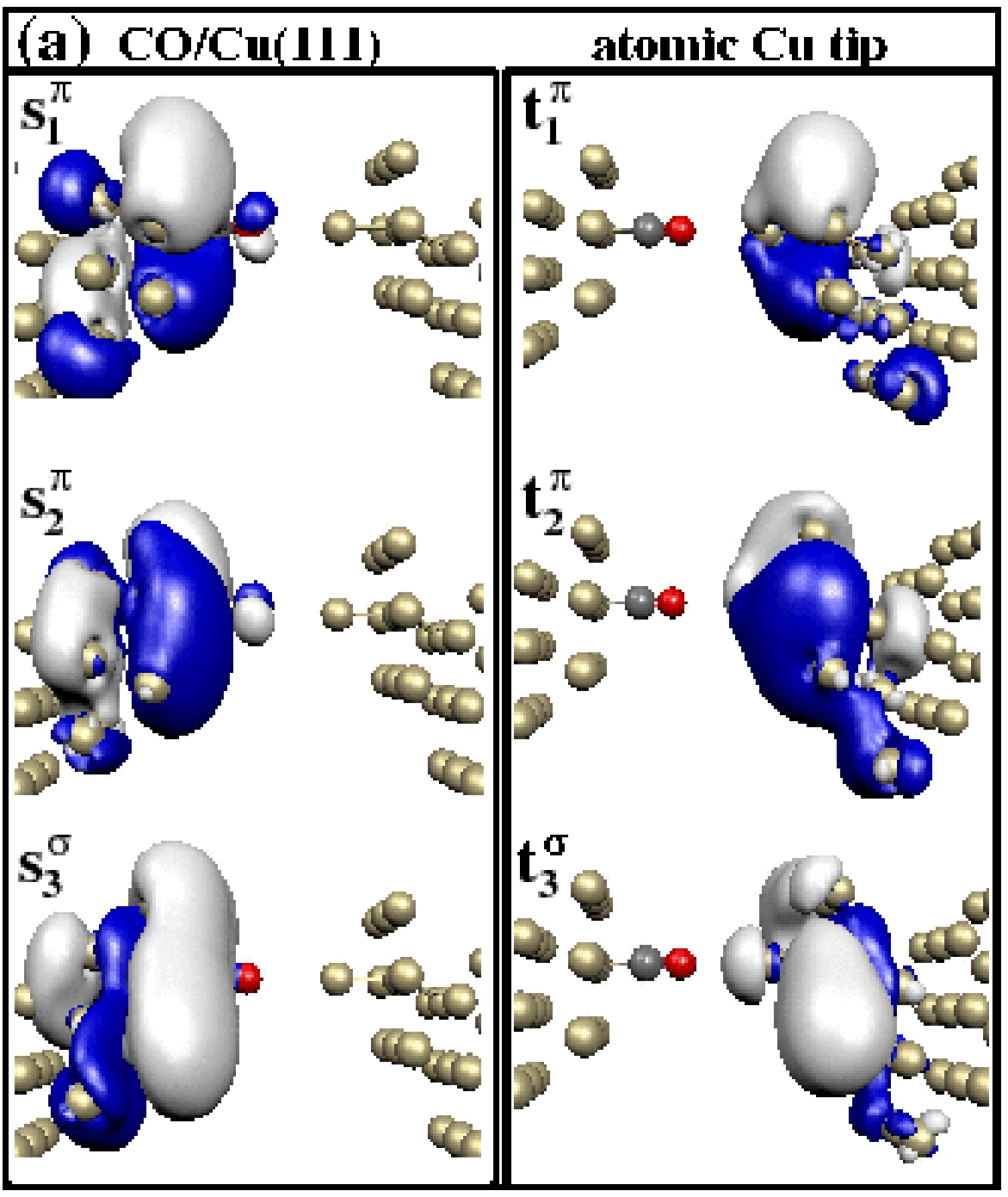}
& \includegraphics[height=6.5cm]
{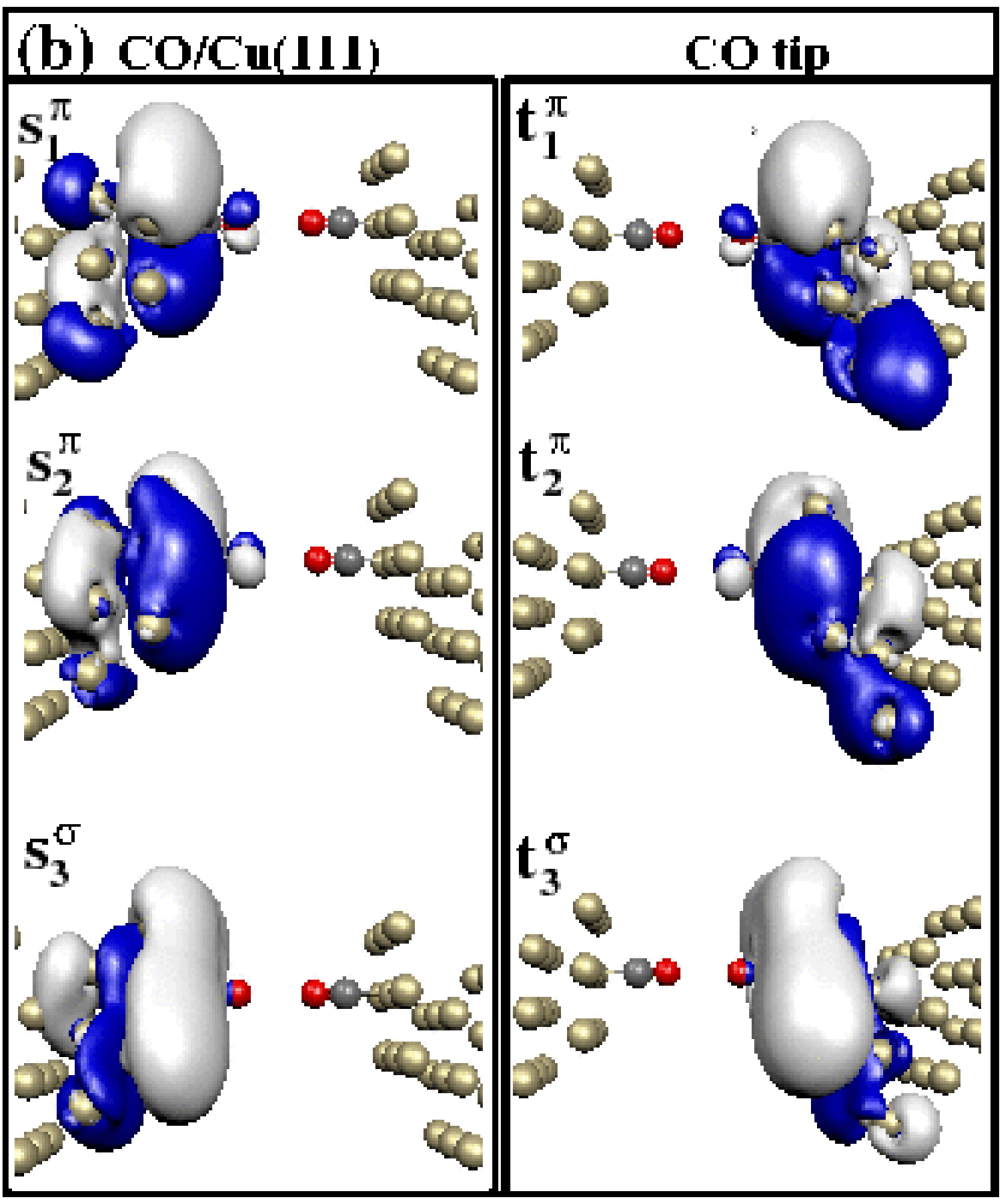}
\\
\includegraphics[height=6.5cm]
{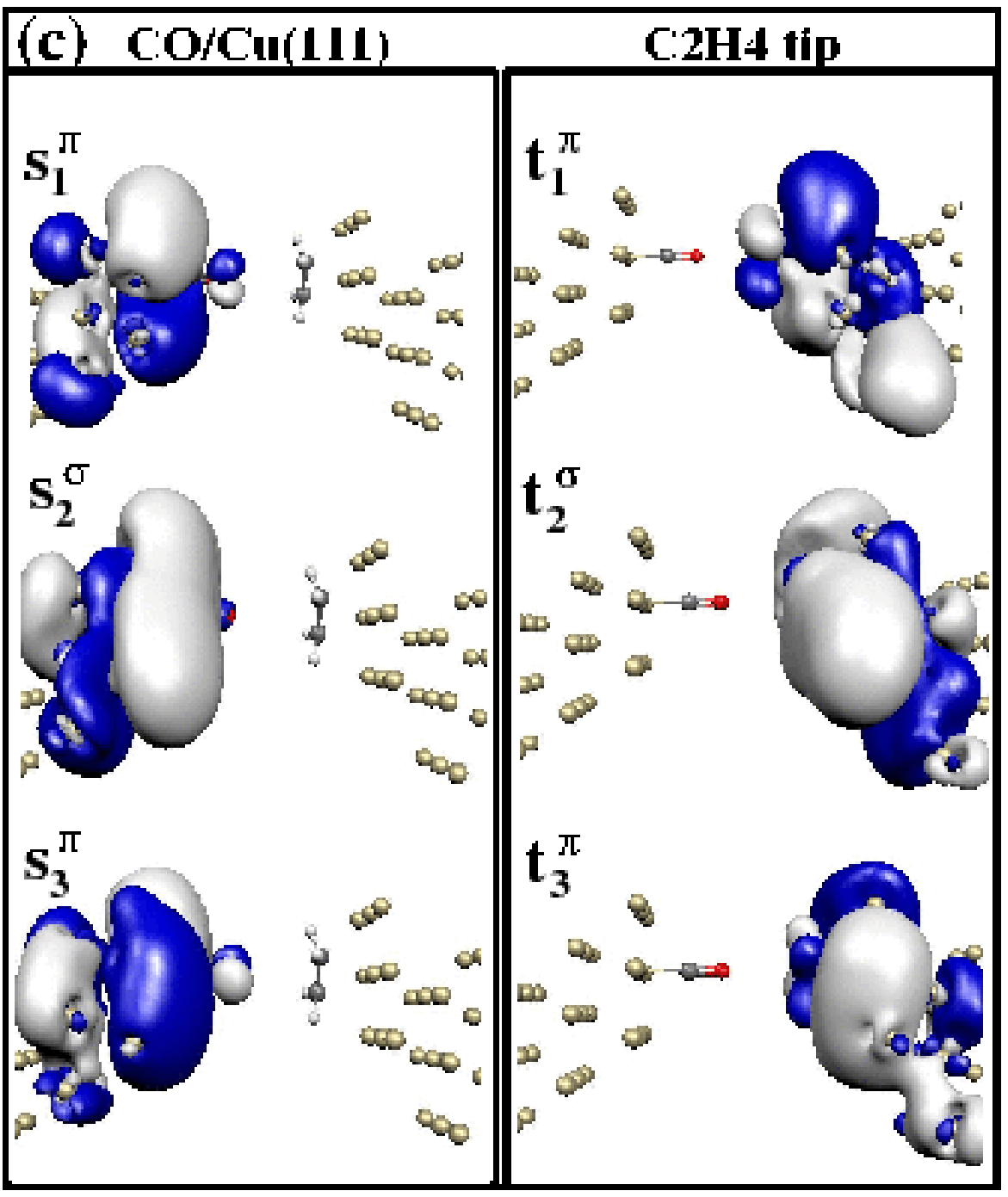}
&
\includegraphics[height=6.5cm]
{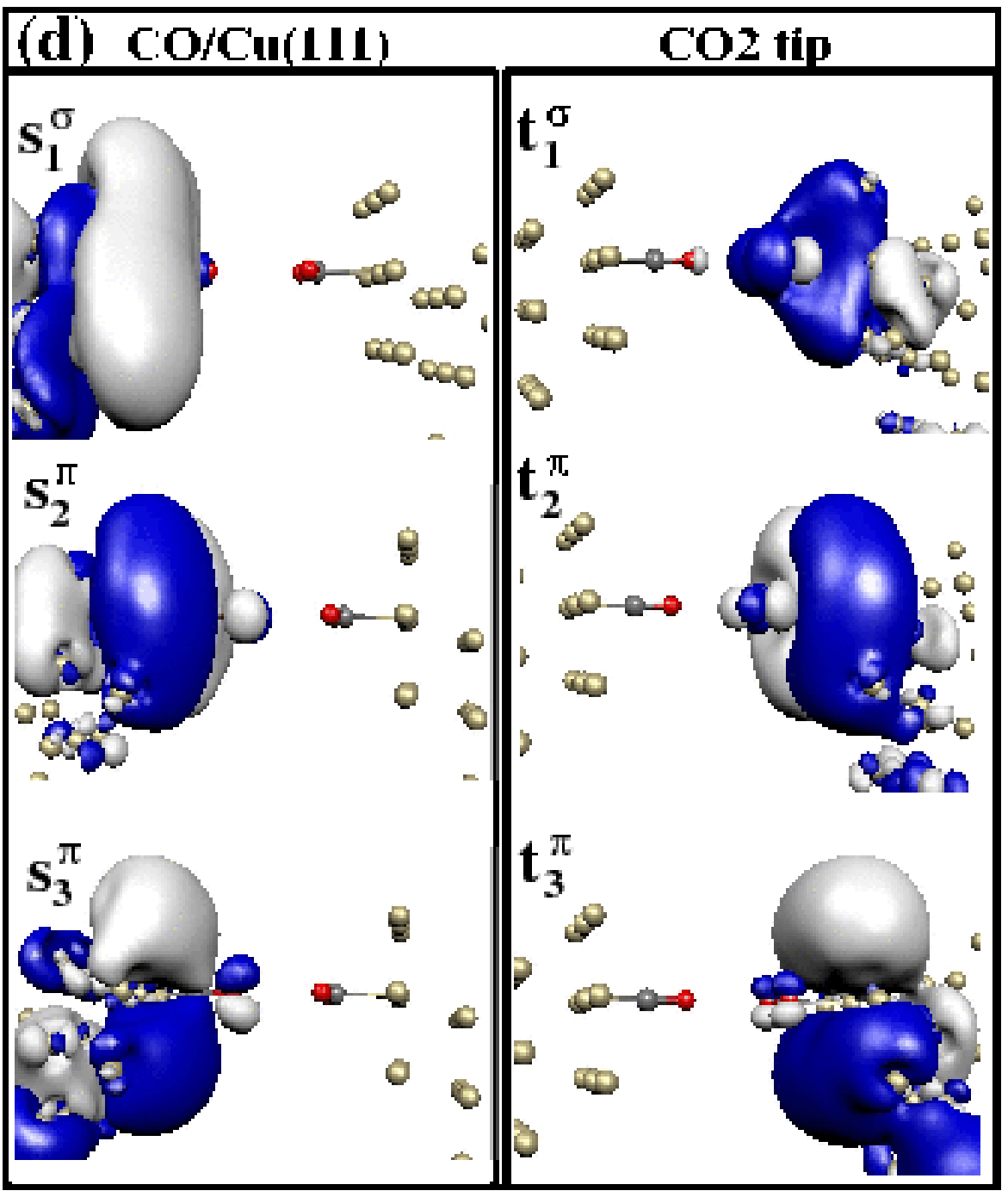}
\\
\multicolumn{2}{c}{\includegraphics[width=5.5cm]
{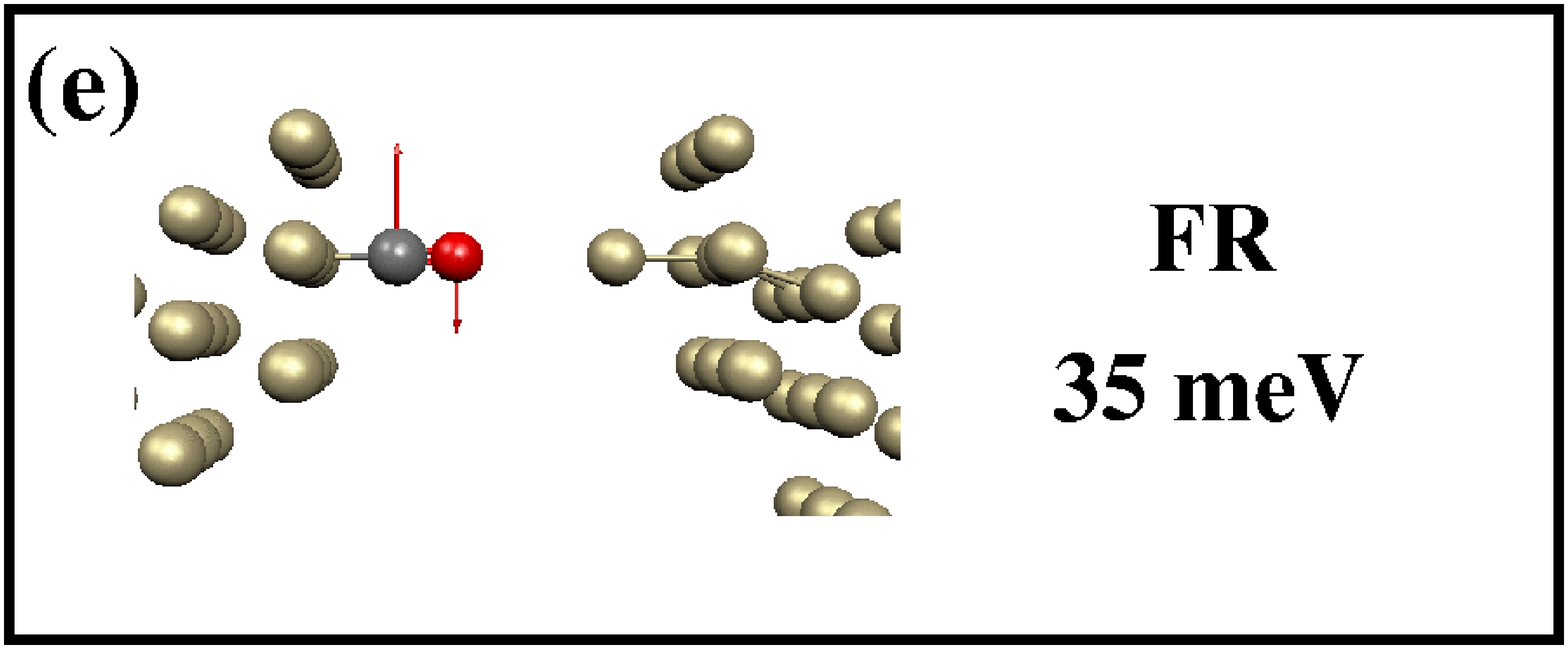}}
\end{tabular}}
\begin{center}
\parbox{15cm}{
\caption{\label{fig2}
Isosurface plots of the three most transmitting
eigenchannels
for each of the four tips considered: (a) Cu adatom tip, (b) CO tip, (c) C$_2$H$_4$ tip, and (d) CO$_2$ tip.
The states labeled $s_i$ ($t_i$) correspond to electron waves incoming from sample/left (tip/right) side. Due to the
tunnel gap between tip and sample the electron scattering states decay rapidly and the transmitted part of the
wave on the other electrode is not visible.
The blue and white colors represent the sign of the real part of 
scattering states (our choice of phase makes the imaginary part 
negligible for visualization purposes).
(e) Visualization of the FR vibrational mode of CO.
}}
\end{center}
\end{figure*}

Three conductance channels dominate the elastic transport through the CO molecule
with different rotational symmetry around the molecular axis.
As shown in \Figref{fig2},
two channels have $\pi$ character and one has $\sigma$ character.\cite{PaBr.07.Transmissioneigenchannelsfrom}
Since electron propagation can occur through each channel in either
direction (from sample to tip or vice versa) there 
are thus two different current-carrying scattering states or wave functions
(denoted $s_i$ and $t_i$) associated with the $i$'th channel.
In the STM setup these scattering states are 
exponentially damped across the tunneling barrier
and retain only a tiny amplitude on the transmitted side (related to the 
transmission coefficient). Among the three dominant sample
states $\{s_1^\pi,s_2^\pi, s_3^\sigma\}$ (i.e., CO side of the junction)
we see in \Figref{fig2} that the weight of the $s^\sigma$ state is strongly reduced
in the molecular apex (O atom) compared with the more extended $s^\pi$
states.
As expected, the symmetry and spatial distribution of the
sample states reflect to some extent
those of the frontier orbitals of CO: the two-fold
degenerate $\pi$-type LUMO, and the $\sigma$-type HOMO
which is more localized on the C atom.

The symmetry of the channels is an important
ingredient to determine the inelastic cross-section
associated with a given vibrational mode.\cite{paulsson_PRL08}
Given the symmetry of the FR mode ($\pi$ character with respect to the CO molecule symmetry axis), 
cf.~\Figref{fig2}(e), we can formulate 
the selection rules in a quite concise form: 
sample states with $\pi$ character
couple only to tip states with $\sigma$ character around the
axis of the sample CO molecule (i.e., $s_i^\pi\leftrightarrow t_j^\sigma$)
and, conversely, sample states with $\sigma$ character
couple only to tip states with $\pi$ character
\revision{(i.e., $s_i^\sigma\leftrightarrow t_j^\pi$)}.\cite{noteA}
Note that here we always define the $\pi$ or $\sigma$ symmetry character of the transmission channels 
with respect to the \revision{axis normal to the surface, i.e., with respect to the molecular axis,} 
and not in the conventional way \revision{for molecular orbitals}. 
Therefore, its meaning might be different, in particular when the two symmetry axes do not coincide.

It is important to note that symmetry
considerations only allow to determine which scattering
events are possible. The relative contribution
of each allowed transition to the total inelastic
cross-section depends on other factors. In particular, 
inelastic scattering rates strongly 
depend on the extension of 
the wave functions of the 
scattering states (both from sample and tip)
along the molecule and towards the 
vacuum. Since we focus on molecular
vibrations, the 
cross section of each scattering event is 
determined by the overlap
between these scattering states within the
gap region, where the molecule is located and
the deformation potential associated with the
vibration is stronger. 
From this simple reasoning, and given their
different spatial distribution, we can expect that
transitions in which $s^\sigma$ of the CO molecule is involved 
will be less efficient that those involving $s^\pi$ channels
of the same molecule. As we will see, this is indeed
confirmed by our calculations.

In the present theoretical work
we analyze the effect of changing the symmetry and extension of
the tip states via functionalization of 
a metallic tip with different molecules, cf.~$t_i$ states in \Figref{fig2}.
Our model tip is composed by a flat Cu(111) surface with 
or without a Cu adatom, or with one of these three different molecules:
CO, CO$_2$, and C$_2$H$_4$.
In a first set of calculations, the 
tip apex is always aligned
with the sample CO molecule,
so both sample and tip share a common symmetry axis. The use of these aligned sample-tip arrangements
allows us to investigate the 
effect of the intrinsic symmetry of the tip states
on the inelastic transport properties using selection rules.
In a second set of calculations, the tips are laterally displaced. 
The main effect
of this displacement is to relax the symmetry constraints
of the system, 
increasing the number of allowed transitions,
and to change the distance between the
sample and tip molecules, thus affecting the overlap
between scattering states coming from both sides.
Although the lateral displacement strongly affects the inelastic transport
in some cases, it is important to note that such changes
can still be completely rationalized
using the same ingredients that allow to understand
the more symmetric situations: (i) 
the local character ($\sigma$ or $\pi$) 
of the most conducting tip states around the sample CO axis, and 
(ii) the overlap between sample 
and tip states with the deformation potential.
All the results reported in the following correspond to the 
tunneling regime, with distances between the  
most protruding tip and sample atoms $\ge 3\, $\AA.

\subsection{Aligned tips}\label{symmetric}

\begin{figure}[t]
\includegraphics[width=8.5cm]{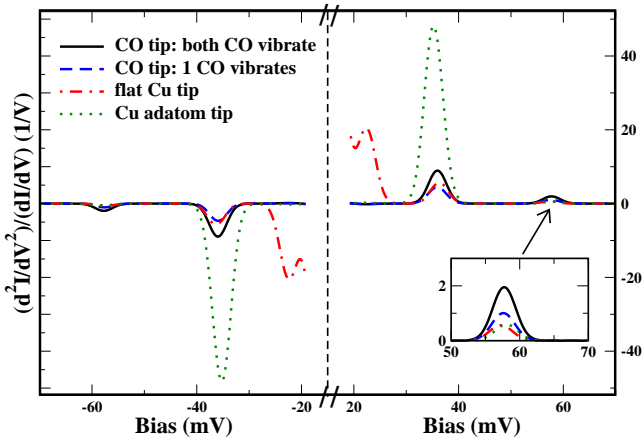}
\vspace{0.5cm}
\caption{\label{fig3}
Symmetrized IETS spectra of a CO
molecule adsorbed on a Cu(111) surface with different models for the tip:
(i) a CO functionalized tip that is either vibrating (solid line)
 or frozen (dashed line),
(ii) a vibrating flat Cu(111) tip (dashed-dotted line) or (iii) a vibrating 
sharp Cu adatom tip (dotted line).
\revision{As a reference, the calculated relative change in tunneling conductance due 
to the frustrated rotation (FR) modes is 24\% with the Cu adatom tip but only
3\% with the flat Cu tip.}
The inset shows a zoom on
the signal related to the CO center-of-mass motion around $57$ mV.}
\end{figure}

\begin{figure}[t]
$\begin{array}{c}
\includegraphics[width=8.5cm]{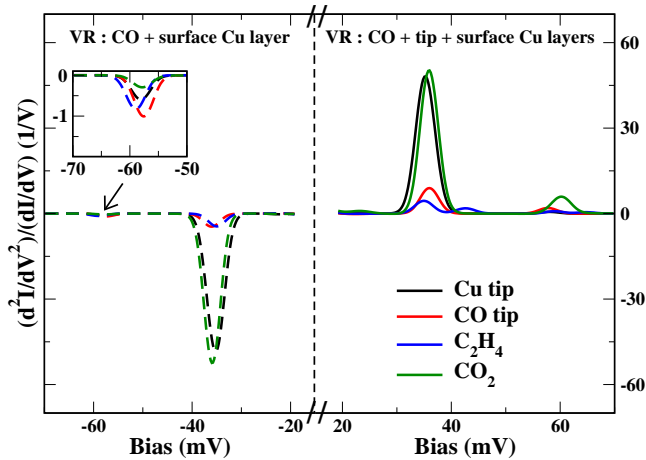} \\ 
\end{array}$
\caption{\label{fig4}
Symmetrized IETS  spectra of a CO
molecule adsorbed on Cu(111) surface using a Cu adatom tip or a tip functionalized with CO,
C$_2$H$_4$, or CO$_2$.
Two vibrational 
regions are considered: (i) top-most Cu layers, sample CO molecule and tip molecule vibrating
(shown for $V>0$), and (ii) only sample side vibrating (shown for $V<0$). 
\revision{As a reference, the calculated relative change in tunneling conductance due 
to the frustrated rotation (FR) modes is 25\% with the CO$_2$ tip but only 2\% with the C$_2$H$_4$ tip.}
The inset shows a zoom
on the signal related to the CO center-of-mass motion around $-57$ mV.}
\end{figure}

Inelastic spectra for CO/Cu(111) using a Cu adatom tip
have recently been experimentally and theoretically
characterized.\cite{paulsson_PRL08,vitali_nano10}
We will use this system
as a standard reference in what follows.
The generic IETS spectrum is dominated by the FR
of the CO molecule
at $\sim 35$\,meV, cf.~\Figref{fig3}. At lower energies, not considered
here, 
the FT of CO gives a
weaker signal at $\sim 5$\,meV. 
Similarly, in our simulations with functionalized tips presented
below we also find the spectra dominated
by the FR of the CO molecule at $\sim 35$\,meV 
(see \Figref{fig4})
for all the 
different tip terminations, albeit with varying intensity.
However, there are also important differences that deserve
a careful analysis.

First, we examine the case of the CO functionalized tip,
where the top-most Cu(111) layers and both tip and sample CO
molecules are allowed to vibrate.
The corresponding theoretical IETS spectrum is shown 
in \Figref{fig3} together with the result for 
the Cu adatom tip. 
In addition to the FR signal, we find a weaker feature that originates in 
the center-of-mass (CM) motion (Cu-CO stretch)  
at $\sim 57$\,meV (see inset to \Figref{fig3}).
Clearly, the use of a CO functionalized tip enhances the CM signal, which 
was hardly observed with a Cu adatom tip.\cite{vitali_nano10}
This effect follows from the   
selection rules described 
in Ref.~\onlinecite{paulsson_PRL08}, and the
observation of such an enhancement in the CM signal might be employed for the 
characterization of the STM tip used in experiments.

In order to check if the presence of the two CO molecules at the
relatively short tunneling distance used here introduces some coupling 
of the vibrations or non-linear effect in the electron-phonon coupling,
we have performed a calculation in which 
one side of the junction is kept frozen. This asymmetric vibrational 
region yields asymmetric features in the IETS spectrum which we
here average out as described above.
As seen in \Figref{fig3}, the FR signal obtained when both tip and sample CO 
molecules vibrate (black line) is larger than the signal obtained with a frozen CO tip
(blue dashed curve)
by a factor of $\sim 2$, showing that the coupling between
the vibrational modes of the two molecules, 
and its effect in the $e$-ph interaction,
is negligible at the considered distances. Therefore,
the deformation potential associated with 
the CO FR mode is not substantially affected by the presence
of neighboring vibrating molecules, and the changes in the
calculated IETS spectra using different functionalized
tips are mainly related to the different symmetry 
and spatial distribution of the
electronic states in each tip. 

Using a CO functionalized tip and a Ag(110) substrate,
Hahn and Ho~\cite{hahn_PRL01} measured
the relative  
intensity of the FR signal for a CO-vacuum-CO tunnel junction
and for a CO-vacuum-Ag(110) tunnel
junction. The inelastic signal was observed to be considerably
larger in the former case. 
In order to check if we can reproduce this behavior,
we have also calculated the IETS spectrum using a flat Cu tip
represented by a Cu(111) surface. 
Figure~\ref{fig3} shows, in agreement with the experiment,
that the corresponding FR signal (red dash-dotted curve) is 
smaller than the signal
obtained with a CO tip (black line).
We note that the calculated signal at low  energies (around $\pm 20$\,mV) 
originates from the vibrational modes of 
the Cu surface and its analysis is out of the scope of 
this work.\cite{vitali_prb2010} 

Next we consider the two other tip terminations shown in \Figref{fig1}. 
As opposed to the CO molecule which is oriented
perpendicular to the Cu(111) surface, the C$_2$H$_4$ and CO$_2$  molecules
form almost planar 
configurations on Cu(111).\cite{witko_ACA98,skibbe_JCP09,wang_SS04}
The corresponding IETS spectra are represented
in \Figref{fig4}, where
the positive voltage region ($V>0$) corresponds to a realistic vibrational region which  includes the sample CO, the 
tip and the top-most Cu layers, while the negative voltage region ($V<0$) shows the results obtained 
for the imaginative but illustrative case when only the sample side (CO molecule and topmost Cu layer) vibrates. A comparison
of the signals shown for these two types of vibrational regions allows us to conclude that the IETS signal
around $\pm 35$\,mV originates from the FR
modes of CO for all four tip models (the signal amplitude is practically unchanged by
freezing the tip atoms).
Furthermore, 
with the CO$_2$ tip frozen the CO CM signal shown around $-60$\,mV is suppressed 
(see inset to \Figref{fig4}). The larger signal observed with the CO$_2$ tip vibrating
(shown for $V>0$) thus originates in tip vibrations.
The FR signal is stronger 
for the CO$_2$ tip than for the Cu adatom tip, which 
in turn is much larger than the signal with CO or C$_2$H$_4$ tips. 
A small FR signal with the C$_2$H$_4$ tip is consistent with
the experimental observations reported in Ref.~\onlinecite{hahn_PRL01}.

Now we discuss our findings using the above mentioned symmetry considerations 
for both the vibrational mode, more precisely the corresponding deformation potential, 
and wave functions of the relevant (tip and sample) scattering states, as well as their 
spatial extension. The inelastic scattering rate for a given phonon mode $\lambda$ 
can be expressed as\cite{paulsson_PRL08}
\begin{equation}\label{inel_rate}
\gamma_{\lambda} = \frac{4\pi\,e}{\hbar}\sum_{i,j}
|\left<s_i|M^\lambda|t_j\right>|^2, \quad |eV|>\hbar\omega_\lambda
\end{equation}
where $M^\lambda$ represents the $e$-ph coupling and the summation
runs over all sample ($s_i$) and tip ($t_i$) scattering states.
As reported in Ref.~\onlinecite{paulsson_PRL08},
in this eigenchannel basis it suffices to include
only the most transmitting eigenchannels for the evaluation
of \Eqref{inel_rate}. This is due to the fact that only these
scattering states have significant spatial weight in the molecular part of the
junction where the $e$-ph interaction (i.e., deformation potential) 
associated with the molecular vibrations is localized.
In \Figref{fig2} we visualize the three most
transmitting eigenchannels as represented by scattering states originating
from the CO sample side ($s_i$) or from the tip side ($t_i$). 
By construction the scattering states belonging to a given eigenchannel 
have the same symmetry. 
As one can see in \Figref{fig2}, 
the three dominant channels are either $\sigma$- or 
$\pi$-type.

\begin{table}[t]
\centering
\begin{ruledtabular}
\begin{tabular}{cccc}
{\bf Cu tip} (99.4) & $t_1^\pi$&$t_2^\pi$&$t_3^\sigma$\\ \hline
$s_1^\pi$ & 0.0   & 0.1   & 46.8   \\
$s_2^\pi$ & 0.0    & 0.1    & 47.8   \\
$s_3^\sigma$ & 2.2    & 2.3    & 0.1  \\
\hline
$\tau_i/10^{-4}$ & $3.74$ &  $3.73$ & $3.33$\\
\hline\hline
{\bf CO tip} (94.1) & $t_1^\pi$&$t_2^\pi$&$t_3^\sigma$\\ \hline
$s_1^\pi$ & 0.0  & 0.0    & 22.7    \\
$s_2^\pi$ & 0.0  & 0.0    & 22.8    \\
$s_3^\sigma$ & 22.8 &22.9    & 0.0   \\
\hline
$\tau_i/10^{-4}$ & $1.02$& $1.02$ & $0.10$ \\
\hline\hline
{\bf C$_2$H$_4$ tip} (80.8) & $t_1^\pi$&$t_2^\sigma$&$t_3^\pi$\\ \hline
$s_1^\pi$ & 0.0   & 0.4     &  0.1     \\
$s_2^\sigma$ & 72.5   &  0.0     & 4.8     \\
$s_3^\pi$ & 0.2    & 2.1     & 0.3   \\
\hline
$\tau_i/10^{-4}$ & $8.18$   & $0.88$  & $0.63$  \\
\hline\hline
{\bf CO$_2$ tip} (99.8) & $t_1^\sigma$&$t_2^\pi$&$t_3^\pi$\\ \hline
$s_1^\sigma$ & 0.0   & 1.9      & 0.1      \\
$s_2^\pi$ & 39.1    & 0.0      &  0.0    \\
$s_3^\pi$ & 58.6     & 0.0      & 0.0    \\
\hline
$\tau_i/10^{-4}$ & $5.68$ & $2.53$  & $0.20$   \\
\end{tabular}
\end{ruledtabular}
\caption{\label{table1}
Decomposition (in \%) of the total inelastic scattering rate for the
frustrated rotations (FR$_{1}$ and FR$_{2}$) of CO on Cu(111)
among the three dominant scattering states only
(i.e., a $3\times 3$ matrix). The sum over all 9 matrix elements,
stated next to the tip label, does not reach 100 \% because of
the total inelastic scattering rate also involves scattering with
the other, less-conducting channels not analyzed in this work.
The elastic transmission coefficient
for each channel  is also given ($\tau_i=\tau_i^s=\tau_i^t$).
}
\end{table}

For each vibrational mode it follows from \Eqref{inel_rate} that
only scattering states with certain
symmetry can couple, giving  rise to the so-called
selection or propensity rules.\cite{paulsson_PRL08}
The two-fold degenerate FR mode corresponds to the rotation of the 
CO molecule parallel to the $xy$ plane. The associated
deformation potential has a $\pi$ character with respect to the CO axis. 
Correspondingly, the FR mode
can only produce scattering of electrons between
$\pi$- and  $\sigma$-type channels. 
This simple picture is confirmed
by our numerical calculations for the 
aligned junctions. 
Table \ref{table1} gives the calculated contributions 
of the transitions between the three most transmitting
eigenchannels
to the total inelastic scattering rate of the two FR modes of CO, 
as obtained 
with different tips.
For the Cu adatom tip the main contribution to the inelastic scattering
comes from the transitions $\{s_1^\pi, s_2^\pi\} \leftrightarrow t_3^\sigma$
and amounts to $\sim 95\%$ of the total scattering rate.
Similarly, for the CO$_2$ tip the main contribution is $\{s_2^\pi, s_3^\pi\} \leftrightarrow t_1^\sigma$ 
($\sim 98\%$). In these two cases the dominant tip state has $\sigma$ character,
while the sample states have $\pi$ character.
On the contrary, in the case of the C$_2$H$_4$ tip the
dominant transition is $s_2^\sigma \leftrightarrow t_1^\pi$ ($\sim73\%$).
For the CO tip, since sample and tip 
subsystems are identical, the dominant 
contribution ($\sim 94\%$) is 
given by the sum of transitions between 
$ \{s_1^\pi, s_2^\pi\} \leftrightarrow t_3^\sigma$ and 
between $s_3^\sigma \leftrightarrow \{t_1^\pi, t_2^\pi\}$. 

In the case of the Cu adatom and CO tips [see \Figref{fig1}(a,b)]
there exists an axial symmetry in the junction due
to their vertical adsorption and  the hexagonal 
symmetry of the substrate. This implies that  
the two $\pi$ eigenchannels and the two FR modes are identical. 
For the Cu adatom tip, the tip state $t_{3}^\sigma$ extends
more towards vacuum than $t_1^\pi$ and $t_2^\pi$, cf.~\Figref{fig2}(a). 
In principle, we can relate the $\sigma$ channel
to the 4$s$ orbital of the Cu adatom, for which 
we can expect to find a considerable spectral weight 
at energies nearby the Fermi energy. According 
to the selection rules, this tip state can couple with the
sample states $\{s_1^\pi,s_2^\pi\}$. Since
$s_1^\pi$ and $s_2^\pi$ also extend the most 
along the molecule and into vacuum, 
the transitions $\{ s_1^\pi,s_2^\pi\} \leftrightarrow t_3^\sigma$ ($\sim 95\%$)
completely dominate the inelastic scattering. 
As expected by symmetry, 
each of these two transitions contribute almost
equally to the total inelastic rate.
For the same reason,
the transitions $s_3^\sigma\leftrightarrow \{t_1^\pi,t_2^\pi\}$ ($\sim 2\%$) also contribute equally, although
in a much smaller percentage due to the smaller
overlap of $s_3^\sigma$ with the vibrational region
and with $\{t_1^\pi,t_2^\pi\}$ of the Cu adatom tip.
With the same axial symmetry, 
the situation is similar in the case of the CO tip, where 
the transitions
$s_3^\sigma \leftrightarrow t_1^\pi$, $s_3^\sigma \leftrightarrow t_2^\pi$,
$s_1^\pi \leftrightarrow t_3^\sigma$, and  $s_2^\pi \leftrightarrow t_3^\sigma$
yield a $\sim 23$\% contribution each, i.e., in this 
case tip and sample states
with both symmetries become relevant.
However, since with two CO
in the junction the $\sigma$-type states are
always rather localized near the metallic surface,
the inelastic signal is significantly reduced with respect to that of the 
Cu adatom tip. This is so, even if we take into account that, 
with two vibrating CO molecules in the junction, the inelastic 
signal associated with the FR modes of CO doubles with respect 
to that of a single vibrating CO molecule. 

Although in this paper we mainly focus on the FR mode of CO,
the reason
for the intensity increase of
the CM mode when a CO tip is used also deserves some discussion.
Contrary to the case of the FR mode,
the deformation potential associated with the CM mode
has  $\sigma$ character around the molecular axis and
induces  transitions
between tip and sample channels of the
same symmetry, i.e., $s_i^\sigma \leftrightarrow t_j^\sigma$
or $s_i^\pi \leftrightarrow t_j^\pi$.
Using a CO tip, the inelastic scattering induced by the
CM mode is dominated by the more extended $\pi$ channels
of CO. Using a Cu adatom tip, the most important
transition induced by the CM mode is $s_3^\sigma \leftrightarrow t_3^\sigma$. However, these
$\sigma$ channels have a much smaller overlap
as compared with the $\pi$ channels that dominate the
case of the CO tip. For this reason, the CO tip
considerably enhances the CM signal.

As for CO$_2$  and C$_2$H$_4$ tips, 
they exhibit in-plane adsorption on Cu(111) and hence,
define a preferential direction given by the molecular axis 
within the $xy$ plane.
Consequently, the two $\pi$ channels are not identical anymore and 
the scattering cross-section of the two FR modes also differs.
In the case of the CO$_2$ tip the most
important processes are the transitions $\{s_2^\pi,s_3^\pi\}\leftrightarrow t_1^\sigma$ ($\sim 98\%$),
i.e., between sample $\pi$ states and tip $\sigma$ states, just as in the case of
 the Cu adatom tip. 
However, 
the contribution from $s_3^\pi \leftrightarrow t_1^\sigma$  is   
larger than the contribution from  $s_2^\pi \leftrightarrow t_1^\sigma$.
This symmetry breaking is also observed for the
$s_1^\sigma \leftrightarrow t_3^\pi$ component, which is almost negligible in comparison
with the $s_1^\sigma \leftrightarrow t_2^\pi$ component.
In the case of C$_2$H$_4$, we find that the transition
$s_2^\sigma\leftrightarrow t_1^\pi$ amounts to $\sim 73$\% of the total inelastic rate
while $s_2^\sigma\leftrightarrow t_3^\pi$ only gives $\sim 5\%$.
This difference is readily understood from \Figref{fig2}(c) where one notes
that $t_1^\pi$ extends more towards the CO sample than the $t_3^\pi$ state,
and hence has the larger overlap with the vibrational region.

All these results provide numerical evidence of the
selection rules governing the inelastic scattering processes.  
Furthermore, as the  $e$-ph coupling is local in space,
even if the symmetry of the orbitals allows for a transition,
the sample and tip states must also extend quite appreciably 
in the transport direction in order to yield a significant inelastic signal. 
Therefore, we have seen that it is the combination of the symmetry \emph{and} spatial extension 
of the tip and sample states that ultimately determines the IETS intensity.
This is indeed
what is reported in \Figref{fig4}, where the FR signal is 
smallest for the C$_2$H$_4$ tip, due to a strong attenuation 
of the $t^\sigma$ state along the molecular axis as compared
to, e.g., $t^\sigma$ for a Cu tip.
On the contrary, for the CO$_2$ tip the $t_1^\sigma$  state
extends deeper into the vibrational region and thus 
allows for a stronger inelastic FR signal.

\subsection{Laterally displaced tips}\label{asymmetric}

\begin{figure*}[t]
\begin{center}
$\begin{array}{c@{\hspace{0.1cm}}c}
\includegraphics[width=6.5cm]{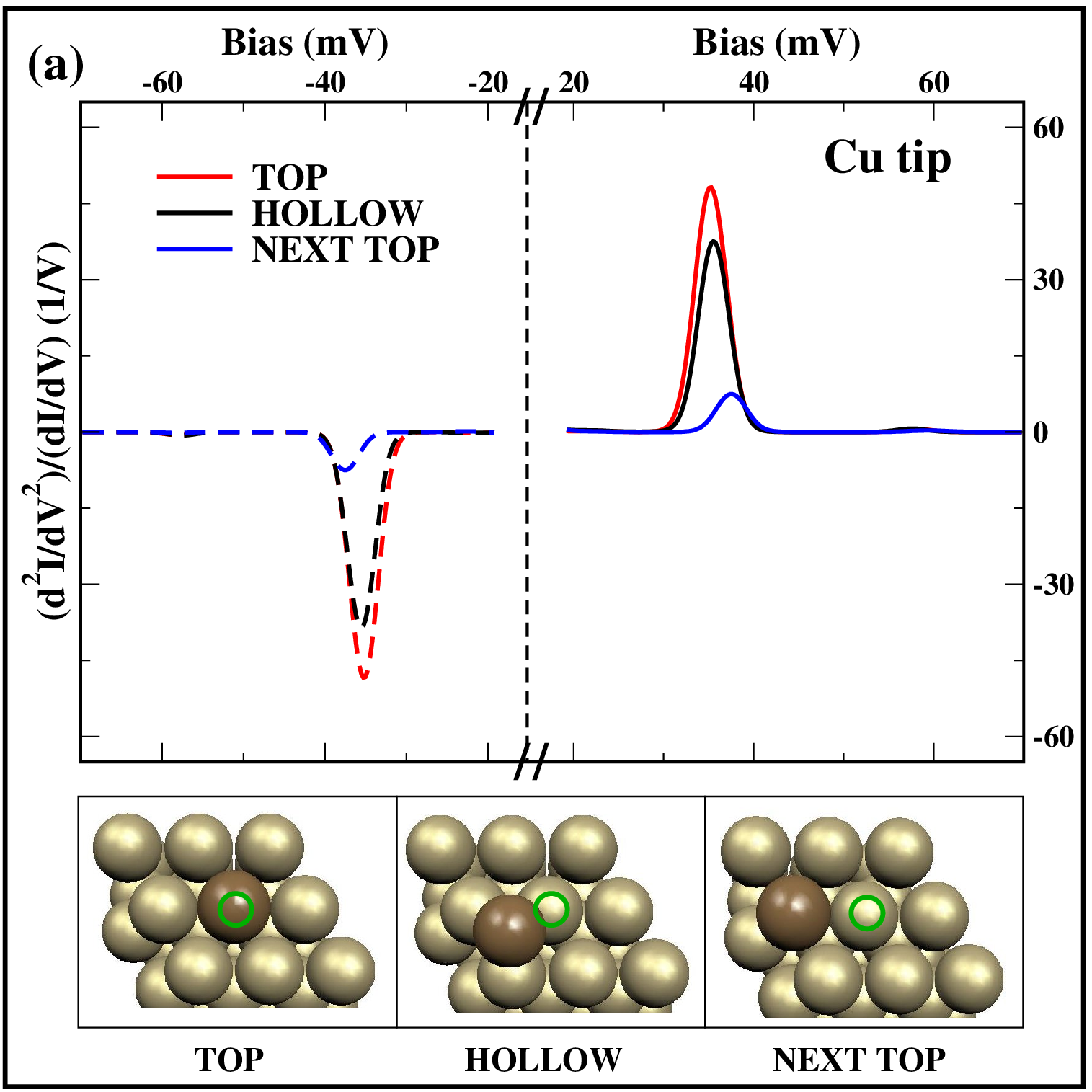} &
\includegraphics[width=6.5cm]{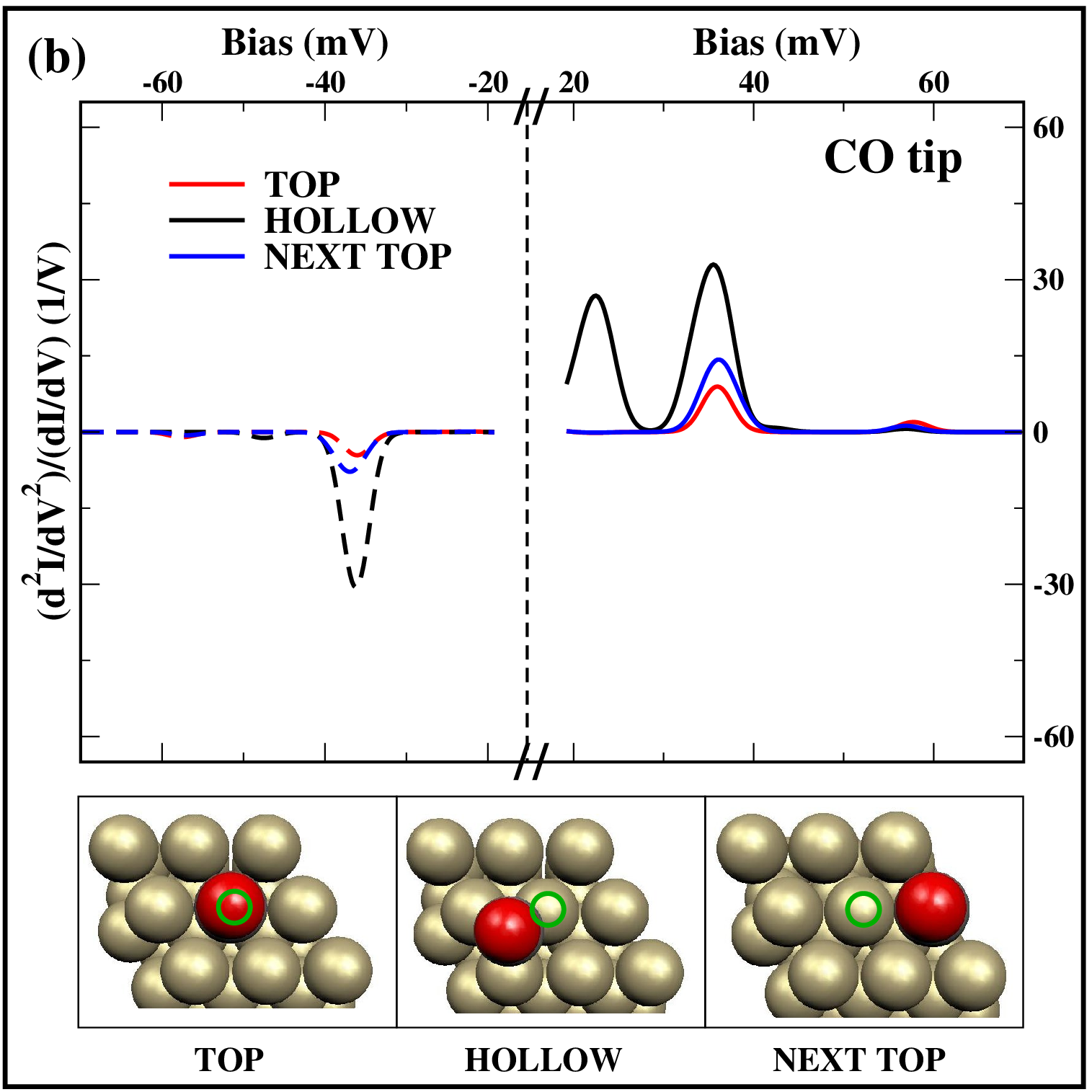} \\ 
\includegraphics[width=6.5cm]{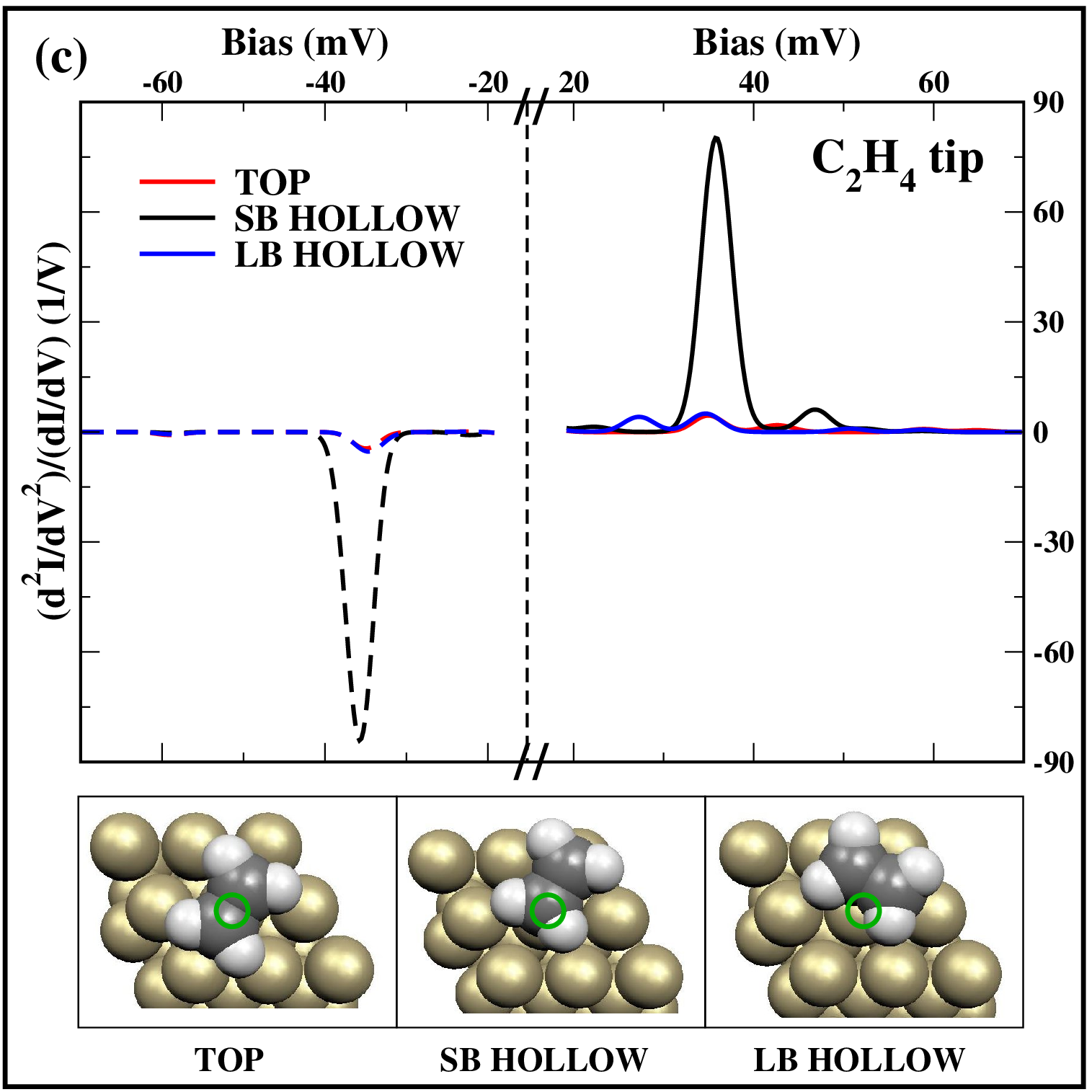} &
\includegraphics[width=6.5cm]{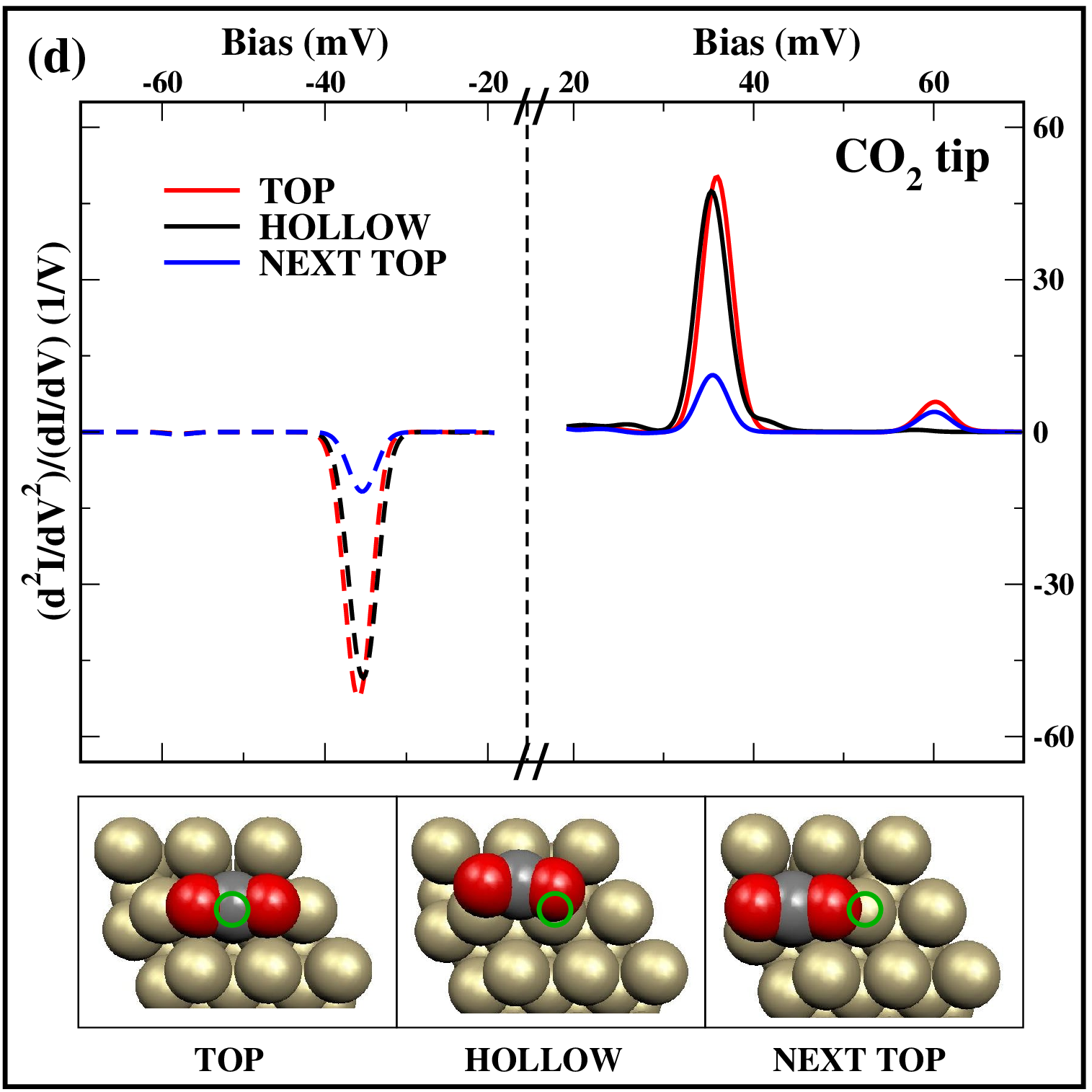} 
\end{array}$
\end{center}
\caption{\label{fig5}
Symmetrized IETS spectra for a CO
molecule adsorbed on Cu(111) surface using (a) a Cu adatom tip, (b) a CO, (c) a  C$_2$H$_4$, or (d) a CO$_2$  tip placed  
on different positions of Cu(111), as shown at the bottom part of each figure. 
Two vibrational regions are considered: (i) top-most Cu layers, CO molecule and tip molecule (shown for $V>0$),
and (ii) only sample CO molecule (shown for $V<0$).
\revision{As a reference, the calculated relative change in tunneling conductance due 
to the frustrated rotation (FR) modes is 24\% with the Cu adatom tip (top site).}
The (green) open circle indicates the position of the sample CO molecule, which always
remains on a Cu top site in the opposite surface.}
\end{figure*}

In the following we 
consider junctions with laterally displaced tips, shown in the lower panels
in \Figref{fig5}. 
If the tip is slightly displaced from the aligned arrangements
considered in \Secref{symmetric} [see \Figref{fig1}], 
the symmetry of the tip-sample system is broken. 
For the Cu adatom or CO tips this means that we no longer have
a common rotation axis in these junction. Similarly, for the 
CO$_2$ and C$_2$H$_4$ tips (which do not carry rotational symmetry)
it means that we reduce the number of symmetry operations.
As illustrated in \Figref{fig5},
we are considering
the following laterally displaced tip-sample configurations:
(i) hollow and next-top configurations for atomic Cu, CO and CO$_2$ tips, and 
(ii) short-bridge and long-bridge hollow configurations for the C$_2$H$_4$ tip.
\revision{We note that the Cu(111) surface and the tip cannot easily be
laterally displaced due to the use of periodic boundary conditions in \textsc{Siesta}.
This would require the use of a substantially larger unit cell (e.g., the use of
two slabs instead of one, or the inclusion of a dislocation in a sufficiently
thick slab, such that the dislocation does not affect the surface electronic
structure). The use of such a large supercell would significantly increase the
computational cost of the calculations. Instead, we have decided here to use a
single  metallic slab. This constraint keeps the computational load at a
moderate level and facilitates the comparison with calculations of the aligned
configuration, although forces us to study only discrete translations of the
functional part of the tip from one adsorption site to the nearest one.}

The results for the corresponding inelastic spectra are also shown 
in \Figref{fig5}. One sees that the intensity of the  
inelastic signal strongly depends on 
the position of the tip. 
In some cases, we also see slight
variations of the energy of the CO FR mode. These small variations
are consistent with those recently observed
in Ref.~\onlinecite{vitali_nano10} when the size of the tunneling
gap is changed. However, in the following we
concentrate only in the reasons behind the change of the intensity of
the inelastic signal. 

In the case of the Cu adatom and the CO$_2$ tips 
[see \Figref{fig5}(a,d)],
the corresponding IETS spectra exhibit almost no change
when displacing the tip from
top to hollow position.
As already shown in \Secref{symmetric},  
the most important tip states for FR scattering with
these two types of tips are the $t^\sigma$ states with $\sigma$
symmetry. 
Therefore, as the tip moves away from the top 
position, the main effect is a modification of  
the  overlap  between a $t^\sigma$-type state
and the deformation potential 
associated with the FR modes.
Such an overlap does not change appreciably as 
the Cu adatom or CO$_2$ tips 
move from top to hollow positions and, as a result, 
the corresponding FR signal is hardly modified. 
However, when the tips are further displaced to the next-top position 
the overlap decreases much more and, consequently, the FR signal is reduced. 
In principle, due to the reduction of symmetry 
the $\sigma$ and $\pi$ classification is only approximate and all
transitions among the scattering states ($s_i\leftrightarrow t_j$)
are allowed. But these contributions are small in practice.
Our calculations suggest that
when a $t^\sigma$ tip state
gives the largest contribution 
to the inelastic signal of the FR mode of CO, this signal 
will decrease as the tip is laterally displaced following the 
reduction of the overlap between 
the electronic states of the tip and sample.

The IETS spectra with CO and C$_2$H$_4$  tips
present a different behavior 
as the tip-sample relative positions  change 
[see \Figref{fig5}(b,c)]. 
The main difference, as compared to the 
previously discussed cases of Cu adatom and CO$_2$ tips, is that,
the FR signal increases when the 
tip is laterally displaced away from the
aligned position. 
The main factor behind this behavior is the increase in the
number of allowed transitions when the symmetry is reduced.
In the case of CO, as the tip is laterally displaced from its 
original top position to hollow site and further to next top,
the FR signal first increases significantly and then 
decreases practically to its initial value. The first increase is
easy to interpret: when the molecule is displaced
inelastic scattering between tip and sample
$\pi$-like states becomes allowed ($\pi \leftrightarrow \pi$). 
Since the $\pi$ states extend the most into the tunneling gap
this transition is more efficient 
than $\sigma \leftrightarrow \pi$, which were the only
allowed transitions for the aligned configuration.
Besides, 
there is a signal at $\sim \pm 20$\,mV
caused by  vibrations of the Cu atoms in the Cu(111) 
surface layers of the sample and the tip, 
similar to the one observed with a flat Cu(111) tip [see \Figref{fig3}].
By further displacing
the tip to the next top position we see a decrease in the FR signal,
consistent with a decrease of the overlap between
tip states with sample states as well as the vibrational region.

\begin{figure}[t]
\begin{center}
\includegraphics[width=8.5cm]{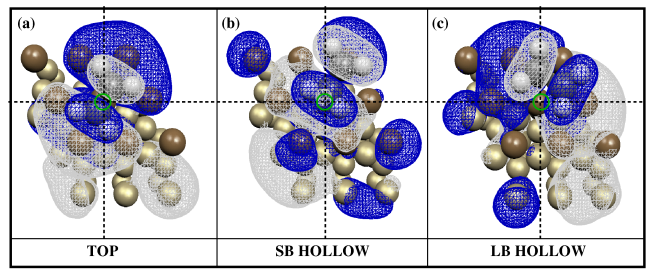} \\
\end{center}
\caption{\label{fig6}
Top view of the tip scattering state which contributes
the most to the inelastic scattering rate.
The green empty circle represents the $xy$ 
position of the sample CO molecule. 
The panels
correspond to the C$_2$H$_4$ molecule placed on (a)
top site, (b) short-bridge hollow, 
and (c) long-bridge hollow positions [as shown in the lower panel  
of \Figref{fig5}(c)]. For clarity, the Cu atoms on the first layer underneath the tip
have been represented in dark brown.
The blue and white colors represent the signs of the real part of 
scattering states (due to the choice of phase the imaginary part is 
negligible for visualization purposes).
}
\end{figure}

A similar behavior is observed in \Figref{fig5}(c) in the case of 
C$_2$H$_4$ functionalization. When the tip molecule is located 
on top site or long-bridge hollow site the FR signal 
is almost identical and small, but
dramatically enhanced if we  move the tip to a short-bridge hollow 
position.
In order to understand the origin 
of this strong enhancement of the FR peak we analyze the spatial
properties of
the dominant tip state ($t_1$) for each of the three C$_2$H$_4$ 
tip positions. 
Figure \ref{fig6} shows a top view  of the three different tip
states, where
the origin (green empty circle) corresponds to the 
position of the sample CO molecule. This allows us to 
survey the ''local'' symmetry 
character of the tip states with respect to the center 
of the CO vibrational mode. As mentioned
in the beginning of \Secref{results}
this is the critical
feature responsible for the amplitude of the IETS signal
of the FR mode.
In Figs.~\ref{fig6}(a,c) we 
observe that for top and long-bridge hollow
tip arrangements, the dominant
tip state has a local $\pi$-type symmetry. For this 
reason the FR mode cannot couple it efficiently to the $\pi$-type
CO sample states, hence the small FR inelastic signal in \Figref{fig5}(c). 
On the contrary,  when the  C$_2$H$_4$ molecule
is placed in the short-bridge hollow position [see \Figref{fig6}(b)]
the dominant eigenchannel has a strong local
$\sigma$-type character and therefore
yields a significantly stronger FR signal.
The IETS peak for the FR modes of the CO is 
even larger than the one observed with a Cu adatom or a CO$_2$ tip on
top positions, due to the fact that the involved
tip states extend even further into the 
vibrational region. 

\section{Conclusions}
\label{sec:conclusions}

In this work the role of the STM tip on inelastic electron tunneling
spectroscopy has been thoroughly investigated. Special emphasis has been placed
on the effects of changes in symmetry of the orbitals involved in the inelastic
scattering processes.

We have simulated IETS spectra for a CO molecule placed on a Cu(111) surface
using a Cu adatom tip as well as tips functionalized with different molecules
(CO, CO$_2$, and C$_2$H$_4$). For all the different tips considered here, the
IETS spectra are dominated by the excitation of the frustrated rotation
vibrational mode of the CO molecule at $\sim$35\,meV. However, we find that the
intensity of the
FR peak strongly depends on the chemical nature as well as on the position of
the STM tip. These effects were rationalized by means of the propensity rules,
i.e., taking into account the local symmetry character of both the deformation
potential associated with the excited vibrational mode and of the most
transmitting scattering states associated with the tip and sample, in the region
where the deformation potential is appreciable.

We performed a detailed analysis of these effects by visualizing the
eigenchannels of the different junctions. In most cases the symmetry and
the spatial distribution of these dominant scattering states can be rationalized
in simple terms and correlated with the frontier molecular orbitals.
Specifically, \revision{our analysis suggests the following consequences of 
IETS propensity rules} related to the symmetry respect to the surface normal:
Adsorbate molecules for which $\pi$-type frontier orbitals control 
the electron transport reveal IETS signals from
transversal (longitudinal) vibrational modes if observed with tips of
$\sigma$-type ($\pi$-type) states. Conversely, adsorbate molecules with
dominating $\sigma$-type frontier orbitals reveal IETS signals from
longitudinal modes if observed with $\sigma$-tips and transverse modes
if observed with $\pi$-tips.
In short, rotational symmetry around the transport
direction favors inelastic transitions where the product of sample state,
deformation potential, and tip state is symmetric.
Note that the $\pi$ or $\sigma$ symmetry character of the transmission channels 
is defined with respect to the \revision{axis normal to the surface} and, therefore, its meaning is not always the 
same as the conventional one of the molecular orbitals.

Our calculations further show that, for a given vibrational mode of the adsorbed
molecule, the inelastic scattering rate is mainly determined by the local
character in the vibrational region of the tip channels and that, as a result,
the relative tip-sample configuration is crucial. Therefore, in the
general cases where sample and tip do not share a symmetry axis, we 
have shown that the propensity rules stated above apply approximately
in terms of the ``local'' symmetry in the region of space where the
molecular vibrations occur.

For the particular case of the frustrated rotation of CO, the locality effect
is specially relevant if the tip states have dominantly $\pi$-type
character, because then the inelastic signal can be largely increased by simply
shifting the tip laterally. We have also found that, in some cases,
functionalized tips can enhance inelastic signals as compared to those obtained
with a metallic adatom tip. For instance, the signal from the center-of-mass
motion of one CO molecule was found to be larger when observed with a CO tip
than with a metallic tip, cf.~inset to \Figref{fig3}.

In conclusion, we have shown that the STM tip is of vital importance in
inelastic tunneling spectroscopy and that, as a result, single-molecule
vibrational spectroscopy could be modulated and manipulated by means of
functionalized STM tips. Moreover, since the effect of the tip might play an
important role in the IETS measurements, it would be desirable to take it into
account in the interpretation of experimental data. We believe our results can
help in that direction.

\begin{acknowledgments}
We thank N.~Lorente, M.~Paulsson, and H.~Ueba for stimulating discussions.
Support from
the Basque Departamento de Educaci\'on, UPV/EHU (Grant No. IT-366-07), 
the Spanish
Ministerio de Ciencia e Innovaci\'on (Grant No. FIS2010-19609-C02-00),
the ETORTEK program funded by the Basque Departamento de Industria and the
Diputacion Foral de Guipuzcoa are gratefully acknowledged.

\end{acknowledgments}


\begin{thebibliography}{50}
\expandafter\ifx\csname natexlab\endcsname\relax\def\natexlab#1{#1}\fi
\expandafter\ifx\csname bibnamefont\endcsname\relax
  \def\bibnamefont#1{#1}\fi
\expandafter\ifx\csname bibfnamefont\endcsname\relax
  \def\bibfnamefont#1{#1}\fi
\expandafter\ifx\csname citenamefont\endcsname\relax
  \def\citenamefont#1{#1}\fi
\expandafter\ifx\csname url\endcsname\relax
  \def\url#1{\texttt{#1}}\fi
\expandafter\ifx\csname urlprefix\endcsname\relax\def\urlprefix{URL }\fi
\providecommand{\bibinfo}[2]{#2}
\providecommand{\eprint}[2][]{\url{#2}}

\bibitem[{\citenamefont{Stipe et~al.}(1998)\citenamefont{Stipe, Rezaei, and
  Ho}}]{stipe_science98}
\bibinfo{author}{\bibfnamefont{B.~C.} \bibnamefont{Stipe}},
  \bibinfo{author}{\bibfnamefont{R.~A.} \bibnamefont{Rezaei}},
  \bibnamefont{and} \bibinfo{author}{\bibfnamefont{W.}~\bibnamefont{Ho}},
  \bibinfo{journal}{Science} \textbf{\bibinfo{volume}{280}},
  \bibinfo{pages}{1732} (\bibinfo{year}{1998}).

\bibitem[{\citenamefont{Jaklevic and
  Lambe}(1966)}]{JaLa.66.MolecularVibrationSpectra}
\bibinfo{author}{\bibfnamefont{R.~C.} \bibnamefont{Jaklevic}} \bibnamefont{and}
  \bibinfo{author}{\bibfnamefont{J.}~\bibnamefont{Lambe}},
  \bibinfo{journal}{Phys.~Rev.~Lett.} \textbf{\bibinfo{volume}{17}},
  \bibinfo{pages}{1139} (\bibinfo{year}{1966}).

\bibitem[{\citenamefont{Komeda et~al.}(2002)\citenamefont{Komeda, Kim, Kawai,
  Persson, and Ueba}}]{KoKiKa.02.Lateralhoppingof}
\bibinfo{author}{\bibfnamefont{T.}~\bibnamefont{Komeda}},
  \bibinfo{author}{\bibfnamefont{Y.}~\bibnamefont{Kim}},
  \bibinfo{author}{\bibfnamefont{M.}~\bibnamefont{Kawai}},
  \bibinfo{author}{\bibfnamefont{B.~N.~J.} \bibnamefont{Persson}},
  \bibnamefont{and} \bibinfo{author}{\bibfnamefont{H.}~\bibnamefont{Ueba}},
  \bibinfo{journal}{Science} \textbf{\bibinfo{volume}{295}},
  \bibinfo{pages}{2055} (\bibinfo{year}{2002}).

\bibitem[{\citenamefont{Heinrich et~al.}(2002)\citenamefont{Heinrich, Lutz,
  Gupta, and Eigler}}]{HeLuGu.2002.MoleculeCascades}
\bibinfo{author}{\bibfnamefont{A.~J.} \bibnamefont{Heinrich}},
  \bibinfo{author}{\bibfnamefont{C.~P.} \bibnamefont{Lutz}},
  \bibinfo{author}{\bibfnamefont{J.~A.} \bibnamefont{Gupta}}, \bibnamefont{and}
  \bibinfo{author}{\bibfnamefont{D.~M.} \bibnamefont{Eigler}},
  \bibinfo{journal}{Science} \textbf{\bibinfo{volume}{298}},
  \bibinfo{pages}{1381} (\bibinfo{year}{2002}).

\bibitem[{\citenamefont{Pascual et~al.}(2003)\citenamefont{Pascual, Lorente,
  Song, Conrad, and Rust}}]{PaLoSo.03.Selectivityinvibrationally}
\bibinfo{author}{\bibfnamefont{J.~I.} \bibnamefont{Pascual}},
  \bibinfo{author}{\bibfnamefont{N.}~\bibnamefont{Lorente}},
  \bibinfo{author}{\bibfnamefont{Z.}~\bibnamefont{Song}},
  \bibinfo{author}{\bibfnamefont{H.}~\bibnamefont{Conrad}}, \bibnamefont{and}
  \bibinfo{author}{\bibfnamefont{H.~P.} \bibnamefont{Rust}},
  \bibinfo{journal}{Nature (London)} \textbf{\bibinfo{volume}{423}},
  \bibinfo{pages}{525} (\bibinfo{year}{2003}).

\bibitem[{\citenamefont{Lorente et~al.}(2005)\citenamefont{Lorente, Rurali, and
  Tang}}]{LoRuTa.05.Single-moleculemanipulationand}
\bibinfo{author}{\bibfnamefont{N.}~\bibnamefont{Lorente}},
  \bibinfo{author}{\bibfnamefont{R.}~\bibnamefont{Rurali}}, \bibnamefont{and}
  \bibinfo{author}{\bibfnamefont{H.}~\bibnamefont{Tang}},
  \bibinfo{journal}{J.~Phys.:~Condens.~Matt.} \textbf{\bibinfo{volume}{17}},
  \bibinfo{pages}{S1049} (\bibinfo{year}{2005}).

\bibitem[{\citenamefont{Ho}(2002)}]{Ho.02.Single-moleculechemistry}
\bibinfo{author}{\bibfnamefont{W.}~\bibnamefont{Ho}},
  \bibinfo{journal}{J.~Chem.~Phys.} \textbf{\bibinfo{volume}{117}},
  \bibinfo{pages}{11033} (\bibinfo{year}{2002}).

\bibitem[{\citenamefont{Appelbaum and Brinkman}(1969)}]{appelbaum_PRB69}
\bibinfo{author}{\bibfnamefont{J.}~\bibnamefont{Appelbaum}} \bibnamefont{and}
  \bibinfo{author}{\bibfnamefont{W.}~\bibnamefont{Brinkman}},
  \bibinfo{journal}{Phys. Rev.} \textbf{\bibinfo{volume}{186}},
  \bibinfo{pages}{464} (\bibinfo{year}{1969}).

\bibitem[{\citenamefont{Caroli et~al.}(1972)\citenamefont{Caroli, Combescot,
  Nozieres, and Saint-James}}]{caroli_JPC72}
\bibinfo{author}{\bibfnamefont{C.}~\bibnamefont{Caroli}},
  \bibinfo{author}{\bibfnamefont{R.}~\bibnamefont{Combescot}},
  \bibinfo{author}{\bibfnamefont{P.}~\bibnamefont{Nozi\`eres}},
  \bibnamefont{and}
  \bibinfo{author}{\bibfnamefont{D.}~\bibnamefont{Saint-James}},
  \bibinfo{journal}{J. Phys. C} \textbf{\bibinfo{volume}{5}},
  \bibinfo{pages}{21} (\bibinfo{year}{1972}).

\bibitem[{\citenamefont{Persson and
  Baratoff}(1987)}]{PEBA.87.INELASTICELECTRON-TUNNELINGFROM}
\bibinfo{author}{\bibfnamefont{B.~N.~J.} \bibnamefont{Persson}}
  \bibnamefont{and} \bibinfo{author}{\bibfnamefont{A.}~\bibnamefont{Baratoff}},
  \bibinfo{journal}{Phys.~Rev.~Lett.} \textbf{\bibinfo{volume}{59}},
  \bibinfo{pages}{339} (\bibinfo{year}{1987}).

\bibitem[{\citenamefont{B\u{o}nca and Trugman}(1995)}]{bonca_PRL95}
\bibinfo{author}{\bibfnamefont{J.}~\bibnamefont{B\u{o}nca}} \bibnamefont{and}
  \bibinfo{author}{\bibfnamefont{S.~A.} \bibnamefont{Trugman}},
  \bibinfo{journal}{Phys. Rev. Lett.} \textbf{\bibinfo{volume}{75}},
  \bibinfo{pages}{2566} (\bibinfo{year}{1995}).

\bibitem[{\citenamefont{Emberly and Kirczenow}(2000)}]{emberly_PRB00}
\bibinfo{author}{\bibfnamefont{E.~G.} \bibnamefont{Emberly}} \bibnamefont{and}
  \bibinfo{author}{\bibfnamefont{G.}~\bibnamefont{Kirczenow}},
  \bibinfo{journal}{Phys. Rev. B} \textbf{\bibinfo{volume}{61}},
  \bibinfo{pages}{5740} (\bibinfo{year}{2000}).

\bibitem[{\citenamefont{Galperin et~al.}(2004)\citenamefont{Galperin, Ratner,
  and Nitzan}}]{GaRaNi.04.Inelasticelectrontunneling}
\bibinfo{author}{\bibfnamefont{M.}~\bibnamefont{Galperin}},
  \bibinfo{author}{\bibfnamefont{M.~A.} \bibnamefont{Ratner}},
  \bibnamefont{and} \bibinfo{author}{\bibfnamefont{A.}~\bibnamefont{Nitzan}},
  \bibinfo{journal}{J.~Chem.~Phys.} \textbf{\bibinfo{volume}{121}},
  \bibinfo{pages}{11965} (\bibinfo{year}{2004}).

\bibitem[{\citenamefont{Mingo and Makoshi}(2000)}]{mingo_PRL00}
\bibinfo{author}{\bibfnamefont{N.}~\bibnamefont{Mingo}} \bibnamefont{and}
  \bibinfo{author}{\bibfnamefont{K.}~\bibnamefont{Makoshi}},
  \bibinfo{journal}{Phys. Rev. Lett.} \textbf{\bibinfo{volume}{84}},
  \bibinfo{pages}{3694} (\bibinfo{year}{2000}).

\bibitem[{\citenamefont{Lorente and Persson}(2000)}]{lorente_PRL00}
\bibinfo{author}{\bibfnamefont{N.}~\bibnamefont{Lorente}} \bibnamefont{and}
  \bibinfo{author}{\bibfnamefont{M.}~\bibnamefont{Persson}},
  \bibinfo{journal}{Phys. Rev. Lett.} \textbf{\bibinfo{volume}{85}},
  \bibinfo{pages}{2997} (\bibinfo{year}{2000}).

\bibitem[{\citenamefont{Frederiksen et~al.}(2004)\citenamefont{Frederiksen,
  Brandbyge, Lorente, and Jauho}}]{FrBrLo.04.InelasticScatteringand}
\bibinfo{author}{\bibfnamefont{T.}~\bibnamefont{Frederiksen}},
  \bibinfo{author}{\bibfnamefont{M.}~\bibnamefont{Brandbyge}},
  \bibinfo{author}{\bibfnamefont{N.}~\bibnamefont{Lorente}}, \bibnamefont{and}
  \bibinfo{author}{\bibfnamefont{A.-P.} \bibnamefont{Jauho}},
  \bibinfo{journal}{Phys.~Rev.~Lett.} \textbf{\bibinfo{volume}{93}},
  \bibinfo{pages}{256601} (\bibinfo{year}{2004}).

\bibitem[{\citenamefont{Alducin et~al.}(2010)\citenamefont{Alducin,
  S\'anchez-Portal, Arnau, and Lorente}}]{AlSaAr.10.Mixed-ValencySignaturein}
\bibinfo{author}{\bibfnamefont{M.}~\bibnamefont{Alducin}},
  \bibinfo{author}{\bibfnamefont{D.}~\bibnamefont{S\'anchez-Portal}},
  \bibinfo{author}{\bibfnamefont{A.}~\bibnamefont{Arnau}}, \bibnamefont{and}
  \bibinfo{author}{\bibfnamefont{N.}~\bibnamefont{Lorente}},
  \bibinfo{journal}{Phys.~Rev.~Lett.} \textbf{\bibinfo{volume}{104}},
  \bibinfo{pages}{136101} (\bibinfo{year}{2010}).

\bibitem[{\citenamefont{Vitali et~al.}(2010{\natexlab{a}})\citenamefont{Vitali,
  Ohmann, Kern, Garcia-Lekue, Frederiksen, Sanchez-Portal, and
  Arnau}}]{vitali_nano10}
\bibinfo{author}{\bibfnamefont{L.}~\bibnamefont{Vitali}},
  \bibinfo{author}{\bibfnamefont{R.}~\bibnamefont{Ohmann}},
  \bibinfo{author}{\bibfnamefont{K.}~\bibnamefont{Kern}},
  \bibinfo{author}{\bibfnamefont{A.}~\bibnamefont{Garcia-Lekue}},
  \bibinfo{author}{\bibfnamefont{T.}~\bibnamefont{Frederiksen}},
  \bibinfo{author}{\bibfnamefont{D.}~\bibnamefont{Sanchez-Portal}},
  \bibnamefont{and} \bibinfo{author}{\bibfnamefont{A.}~\bibnamefont{Arnau}},
  \bibinfo{journal}{Nano Lett.} \textbf{\bibinfo{volume}{10}},
  \bibinfo{pages}{657} (\bibinfo{year}{2010}{\natexlab{a}}).

\bibitem[{\citenamefont{Okabayashi et~al.}(2010)\citenamefont{Okabayashi,
  Paulsson, Ueba, Konda, and
  Komeda}}]{OkPaUe.10.InelasticTunnelingSpectroscopy}
\bibinfo{author}{\bibfnamefont{N.}~\bibnamefont{Okabayashi}},
  \bibinfo{author}{\bibfnamefont{M.}~\bibnamefont{Paulsson}},
  \bibinfo{author}{\bibfnamefont{H.}~\bibnamefont{Ueba}},
  \bibinfo{author}{\bibfnamefont{Y.}~\bibnamefont{Konda}}, \bibnamefont{and}
  \bibinfo{author}{\bibfnamefont{T.}~\bibnamefont{Komeda}},
  \bibinfo{journal}{Phys.~Rev.~Lett.} \textbf{\bibinfo{volume}{104}},
  \bibinfo{pages}{077801} (\bibinfo{year}{2010}).

\bibitem[{\citenamefont{Arroyo et~al.}(2010)\citenamefont{Arroyo, Frederiksen,
  Rubio-Bollinger, V\'elez, Arnau, S\'anchez-Portal, and
  Agra\"it}}]{ArFrRu.2010.Characterizationofsingle-molecule}
\bibinfo{author}{\bibfnamefont{C.~R.} \bibnamefont{Arroyo}},
  \bibinfo{author}{\bibfnamefont{T.}~\bibnamefont{Frederiksen}},
  \bibinfo{author}{\bibfnamefont{G.}~\bibnamefont{Rubio-Bollinger}},
  \bibinfo{author}{\bibfnamefont{M.}~\bibnamefont{V\'elez}},
  \bibinfo{author}{\bibfnamefont{A.}~\bibnamefont{Arnau}},
  \bibinfo{author}{\bibfnamefont{D.}~\bibnamefont{S\'anchez-Portal}},
  \bibnamefont{and} \bibinfo{author}{\bibfnamefont{N.}~\bibnamefont{Agra\"it}},
  \bibinfo{journal}{Phys.~Rev.~B} \textbf{\bibinfo{volume}{81}},
  \bibinfo{pages}{075405} (\bibinfo{year}{2010}).
  
\bibitem[{\citenamefont{Lorente et~al.}(2001)\citenamefont{Lorente, Persson,
  Lauhon, and Ho}}]{lorente_PRL01}
\bibinfo{author}{\bibfnamefont{N.}~\bibnamefont{Lorente}},
  \bibinfo{author}{\bibfnamefont{M.}~\bibnamefont{Persson}},
  \bibinfo{author}{\bibfnamefont{L.~J.} \bibnamefont{Lauhon}},
  \bibnamefont{and} \bibinfo{author}{\bibfnamefont{W.}~\bibnamefont{Ho}},
  \bibinfo{journal}{Phys. Rev. Lett.} \textbf{\bibinfo{volume}{86}},
  \bibinfo{pages}{2593} (\bibinfo{year}{2001}).

\bibitem[{\citenamefont{Troisi and
  Ratner}(2006)}]{TrRa.06.Moleculartransportjunctions}
\bibinfo{author}{\bibfnamefont{A.}~\bibnamefont{Troisi}} \bibnamefont{and}
  \bibinfo{author}{\bibfnamefont{M.~A.} \bibnamefont{Ratner}},
  \bibinfo{journal}{Nano Lett.} \textbf{\bibinfo{volume}{6}},
  \bibinfo{pages}{1784} (\bibinfo{year}{2006}).

\bibitem[{\citenamefont{Gagliardi et~al.}(2007)\citenamefont{Gagliardi,
  Solomon, Pecchia, Frauenheim, Di~Carlo, Hush, and
  Reimers}}]{GaSoPe.07.priorimethodpropensity}
\bibinfo{author}{\bibfnamefont{A.}~\bibnamefont{Gagliardi}},
  \bibinfo{author}{\bibfnamefont{G.~C.} \bibnamefont{Solomon}},
  \bibinfo{author}{\bibfnamefont{A.}~\bibnamefont{Pecchia}},
  \bibinfo{author}{\bibfnamefont{T.}~\bibnamefont{Frauenheim}},
  \bibinfo{author}{\bibfnamefont{A.}~\bibnamefont{Di~Carlo}},
  \bibinfo{author}{\bibfnamefont{N.~S.} \bibnamefont{Hush}}, \bibnamefont{and}
  \bibinfo{author}{\bibfnamefont{J.~R.} \bibnamefont{Reimers}},
  \bibinfo{journal}{Phys.~Rev.~B} \textbf{\bibinfo{volume}{75}},
  \bibinfo{pages}{174306} (\bibinfo{year}{2007}).

\bibitem[{\citenamefont{Paulsson et~al.}(2008)\citenamefont{Paulsson,
  Frederiksen, Ueba, Lorente, and Brandbyge}}]{paulsson_PRL08}
\bibinfo{author}{\bibfnamefont{M.}~\bibnamefont{Paulsson}},
  \bibinfo{author}{\bibfnamefont{T.}~\bibnamefont{Frederiksen}},
  \bibinfo{author}{\bibfnamefont{H.}~\bibnamefont{Ueba}},
  \bibinfo{author}{\bibfnamefont{N.}~\bibnamefont{Lorente}}, \bibnamefont{and}
  \bibinfo{author}{\bibfnamefont{M.}~\bibnamefont{Brandbyge}},
  \bibinfo{journal}{Phys. Rev. Lett.} \textbf{\bibinfo{volume}{100}},
  \bibinfo{pages}{226604} (\bibinfo{year}{2008}).

\bibitem[{\citenamefont{Bartels et~al.}(1999)\citenamefont{Bartels, Meyer, and
  Rieder}}]{BaMeRi.99.evolutionofCO}
\bibinfo{author}{\bibfnamefont{L.}~\bibnamefont{Bartels}},
  \bibinfo{author}{\bibfnamefont{G.}~\bibnamefont{Meyer}}, \bibnamefont{and}
  \bibinfo{author}{\bibfnamefont{K.~H.} \bibnamefont{Rieder}},
  \bibinfo{journal}{Surface Science} \textbf{\bibinfo{volume}{432}},
  \bibinfo{pages}{L621} (\bibinfo{year}{1999}).

\bibitem[{\citenamefont{Teobaldi et~al.}(2007)\citenamefont{Teobaldi, Penalba,
  Arnau, Lorente, and Hofer}}]{TePeAr.07.Includingprobetip}
\bibinfo{author}{\bibfnamefont{G.}~\bibnamefont{Teobaldi}},
  \bibinfo{author}{\bibfnamefont{M.}~\bibnamefont{Pe\~nalba}},
  \bibinfo{author}{\bibfnamefont{A.}~\bibnamefont{Arnau}},
  \bibinfo{author}{\bibfnamefont{N.}~\bibnamefont{Lorente}}, \bibnamefont{and}
  \bibinfo{author}{\bibfnamefont{W.~A.} \bibnamefont{Hofer}},
  \bibinfo{journal}{Phys.~Rev.~B} \textbf{\bibinfo{volume}{76}},
  \bibinfo{eid}{235407} (\bibinfo{year}{2007}).

\bibitem[{\citenamefont{Calleja et~al.}(2004)\citenamefont{Calleja, Arnau,
  Hinarejos, {Vazquez de Parga}, Hofer, Echenique, and Miranda}}]{Calleja10}
\bibinfo{author}{\bibfnamefont{F.}~\bibnamefont{Calleja}},
  \bibinfo{author}{\bibfnamefont{A.}~\bibnamefont{Arnau}},
  \bibinfo{author}{\bibfnamefont{J.~J.} \bibnamefont{Hinarejos}},
  \bibinfo{author}{\bibfnamefont{A.~L.} \bibnamefont{{Vazquez de Parga}}},
  \bibinfo{author}{\bibfnamefont{W.~A.} \bibnamefont{Hofer}},
  \bibinfo{author}{\bibfnamefont{P.~M.} \bibnamefont{Echenique}},
  \bibnamefont{and} \bibinfo{author}{\bibfnamefont{R.}~\bibnamefont{Miranda}},
  \bibinfo{journal}{Phys.~Rev.~Lett.} \textbf{\bibinfo{volume}{92}},
  \bibinfo{pages}{206101} (\bibinfo{year}{2004}).

\bibitem[{\citenamefont{Hahn and Ho}(2001)}]{hahn_PRL01}
\bibinfo{author}{\bibfnamefont{J.~R.} \bibnamefont{Hahn}} \bibnamefont{and}
  \bibinfo{author}{\bibfnamefont{W.}~\bibnamefont{Ho}}, \bibinfo{journal}{Phys.
  Rev. Lett.} \textbf{\bibinfo{volume}{87}}, \bibinfo{pages}{196102}
  (\bibinfo{year}{2001}).

\bibitem[{\citenamefont{Lauhon and
  Ho}(1999)}]{LaHo.1999.Single-moleculevibrationalspectroscopy}
\bibinfo{author}{\bibfnamefont{L.~J.} \bibnamefont{Lauhon}} \bibnamefont{and}
  \bibinfo{author}{\bibfnamefont{W.}~\bibnamefont{Ho}},
  \bibinfo{journal}{Phys.~Rev.~B} \textbf{\bibinfo{volume}{60}},
  \bibinfo{pages}{R8525} (\bibinfo{year}{1999}).

\bibitem[{\citenamefont{Persson}(2004)}]{Pe.04.Theoryofelastic}
\bibinfo{author}{\bibfnamefont{M.}~\bibnamefont{Persson}},
  \bibinfo{journal}{Phil.~Trans.~R.~Soc.~Lond.~A.}
  \textbf{\bibinfo{volume}{362}}, \bibinfo{pages}{1173} (\bibinfo{year}{2004}).

\bibitem[{\citenamefont{Soler et~al.}(2002)\citenamefont{Soler, Artacho, Gale,
  Garcia, Junquera, Ordej\'on, and Sanchez-Portal}}]{siesta}
\bibinfo{author}{\bibfnamefont{J.~M.} \bibnamefont{Soler}},
  \bibinfo{author}{\bibfnamefont{E.}~\bibnamefont{Artacho}},
  \bibinfo{author}{\bibfnamefont{J.~D.} \bibnamefont{Gale}},
  \bibinfo{author}{\bibfnamefont{A.}~\bibnamefont{Garcia}},
  \bibinfo{author}{\bibfnamefont{J.}~\bibnamefont{Junquera}},
  \bibinfo{author}{\bibfnamefont{P.}~\bibnamefont{Ordej\'on}},
  \bibnamefont{and}
  \bibinfo{author}{\bibfnamefont{D.}~\bibnamefont{Sanchez-Portal}},
  \bibinfo{journal}{J. Phys.: Condens. Matter} \textbf{\bibinfo{volume}{14}},
  \bibinfo{pages}{2745} (\bibinfo{year}{2002}).

\bibitem[{\citenamefont{Perdew et~al.}(1996)\citenamefont{Perdew, Burke, and
  Ernzerhof}}]{PeBuEr.96.Generalizedgradientapproximation}
\bibinfo{author}{\bibfnamefont{J.~P.} \bibnamefont{Perdew}},
  \bibinfo{author}{\bibfnamefont{K.}~\bibnamefont{Burke}}, \bibnamefont{and}
  \bibinfo{author}{\bibfnamefont{M.}~\bibnamefont{Ernzerhof}},
  \bibinfo{journal}{Phys.~Rev.~Lett.} \textbf{\bibinfo{volume}{77}},
  \bibinfo{pages}{3865} (\bibinfo{year}{1996}).

\bibitem[{\citenamefont{Brandbyge et~al.}(2002)\citenamefont{Brandbyge, Mozos,
  Ordej{\'o}n, Taylor, and Stokbro}}]{brandbyge_PRB02}
\bibinfo{author}{\bibfnamefont{M.}~\bibnamefont{Brandbyge}},
  \bibinfo{author}{\bibfnamefont{J.~L.}~\bibnamefont{Mozos}},
  \bibinfo{author}{\bibfnamefont{P.}~\bibnamefont{Ordej{\'o}n}},
  \bibinfo{author}{\bibfnamefont{J.}~\bibnamefont{Taylor}}, \bibnamefont{and}
  \bibinfo{author}{\bibfnamefont{K.}~\bibnamefont{Stokbro}},
  \bibinfo{journal}{Phys. Rev. B} \textbf{\bibinfo{volume}{65}},
  \bibinfo{pages}{165401} (\bibinfo{year}{2002}).

\bibitem[{\citenamefont{PaBr07}(2007)}]{PaBr.07.Transmissioneigenchannelsfrom}
\bibinfo{author}{\bibfnamefont{M.} \bibnamefont{Paulsson}}
  \bibnamefont{and}
  \bibinfo{author}{\bibfnamefont{M.}~\bibnamefont{Brandbyge}},
  \bibinfo{journal}{Phys. Rev. B} \textbf{\bibinfo{volume}{76}},
  \bibinfo{pages}{115117} (\bibinfo{year}{2007}).

\bibitem[{Ine()}]{Inelastica}
\bibinfo{note}{\texttt{http://inelastica.sourceforge.net/}}.

\bibitem[{\citenamefont{Frederiksen et~al.}(2007)\citenamefont{Frederiksen,
  Paulsson, Brandbyge, and Jauho}}]{frederiksen_PRB07}
\bibinfo{author}{\bibfnamefont{T.}~\bibnamefont{Frederiksen}},
  \bibinfo{author}{\bibfnamefont{M.}~\bibnamefont{Paulsson}},
  \bibinfo{author}{\bibfnamefont{M.}~\bibnamefont{Brandbyge}},
  \bibnamefont{and} \bibinfo{author}{\bibfnamefont{A.-P.} \bibnamefont{Jauho}},
  \bibinfo{journal}{Phys. Rev. B} \textbf{\bibinfo{volume}{75}},
  \bibinfo{pages}{205413} (\bibinfo{year}{2007}).

\bibitem[{\citenamefont{Paulsson et~al.}(2005)\citenamefont{Paulsson,
  Frederiksen, and Brandbyge}}]{paulsson_PRB05}
\bibinfo{author}{\bibfnamefont{M.}~\bibnamefont{Paulsson}},
  \bibinfo{author}{\bibfnamefont{T.}~\bibnamefont{Frederiksen}},
  \bibnamefont{and}
  \bibinfo{author}{\bibfnamefont{M.}~\bibnamefont{Brandbyge}},
  \bibinfo{journal}{Phys. Rev. B} \textbf{\bibinfo{volume}{72}},
  \bibinfo{pages}{201101(R)} (\bibinfo{year}{2005}).

\bibitem[{\citenamefont{Viljas et~al.}(2005)\citenamefont{Viljas, Cuevas,
  Pauly, and H{\"a}fner}}]{viljas_PRB05}
\bibinfo{author}{\bibfnamefont{J.~K.} \bibnamefont{Viljas}},
  \bibinfo{author}{\bibfnamefont{J.~C.} \bibnamefont{Cuevas}},
  \bibinfo{author}{\bibfnamefont{F.}~\bibnamefont{Pauly}}, \bibnamefont{and}
  \bibinfo{author}{\bibfnamefont{M.}~\bibnamefont{H{\"a}fner}},
  \bibinfo{journal}{Phys. Rev. B} \textbf{\bibinfo{volume}{72}},
  \bibinfo{pages}{245415} (\bibinfo{year}{2005}).

\bibitem[{\citenamefont{Haupt}(2010)}]{HaNoBe.10.Currentnoisein}
\bibinfo{author}{\bibfnamefont{F.} \bibnamefont{Haupt}},
  \bibinfo{author}{\bibfnamefont{T.}~\bibnamefont{Novotn\'y}},
  \bibnamefont{and}
  \bibinfo{author}{\bibfnamefont{W.}~\bibnamefont{Belzig}},
  \bibinfo{journal}{Phys. Rev. B} \textbf{\bibinfo{volume}{82}},
  \bibinfo{pages}{165441} (\bibinfo{year}{2010}).

\bibitem[{\citenamefont{Lopez and N{\o}rskov}(2001)}]{lopez_SS01}
\bibinfo{author}{\bibfnamefont{N.}~\bibnamefont{Lopez}} \bibnamefont{and}
  \bibinfo{author}{\bibfnamefont{J.~K.} \bibnamefont{N{\o}rskov}},
  \bibinfo{journal}{Surf. Sci.} \textbf{\bibinfo{volume}{477}},
  \bibinfo{pages}{59} (\bibinfo{year}{2001}).

\bibitem[{\citenamefont{Alc{\'a}ntara-Ortigoza
  et~al.}(2009)\citenamefont{Alc{\'a}ntara-Ortigoza, Heid, Bohnen, and
  Rahman}}]{alcantara_PRB09}
\bibinfo{author}{\bibfnamefont{M.}~\bibnamefont{Alc{\'a}ntara-Ortigoza}},
  \bibinfo{author}{\bibfnamefont{R.}~\bibnamefont{Heid}},
  \bibinfo{author}{\bibfnamefont{K.-P.} \bibnamefont{Bohnen}},
  \bibnamefont{and} \bibinfo{author}{\bibfnamefont{T.~S.}
  \bibnamefont{Rahman}}, \bibinfo{journal}{Phys. Rev. B}
  \textbf{\bibinfo{volume}{79}}, \bibinfo{pages}{125432}
  (\bibinfo{year}{2009}).

\bibitem[{noteA()}]{noteA}
\bibinfo{note}{Note that inelastic scattering among sample 
or tip states (i.e., $s_i\leftrightarrow s_j$ or $t_i\leftrightarrow t_j$) 
does not yield a bias-dependent signal since their occupations are 
characterized by the same chemical potential. Such processes can thus 
safely be ignored in the discussion of inelastic selection rules.}

\bibitem[{\citenamefont{Vitali et~al.}(2010{\natexlab{b}})\citenamefont{Vitali,
  Borisova, Rusina, Chulkov, and Kern}}]{vitali_prb2010}
\bibinfo{author}{\bibfnamefont{L.}~\bibnamefont{Vitali}},
  \bibinfo{author}{\bibfnamefont{S.~D.}~\bibnamefont{Borisova}},
  \bibinfo{author}{\bibfnamefont{G.~G.}~\bibnamefont{Rusina}},
  \bibinfo{author}{\bibfnamefont{E.~V.}~\bibnamefont{Chulkov}}, \bibnamefont{and}
  \bibinfo{author}{\bibfnamefont{K.}~\bibnamefont{Kern}},
  \bibinfo{journal}{Phys. Rev. B} \textbf{\bibinfo{volume}{81}},
  \bibinfo{pages}{153409} (\bibinfo{year}{2010}{\natexlab{b}}).

\bibitem[{\citenamefont{Witiko and Hermann}(1998)}]{witko_ACA98}
\bibinfo{author}{\bibfnamefont{M.}~\bibnamefont{Witiko}} \bibnamefont{and}
  \bibinfo{author}{\bibfnamefont{K.}~\bibnamefont{Hermann}},
  \bibinfo{journal}{Appl. Catal A: Gen.} \textbf{\bibinfo{volume}{172}},
  \bibinfo{pages}{85} (\bibinfo{year}{1998}).

\bibitem[{\citenamefont{Skibbe et~al.}(2009)\citenamefont{Skibbe, Vogel,
  Binder, Pucci, Kravchuk, Vattuone, Venugopal, Kokalj, and
  Rocca}}]{skibbe_JCP09}
\bibinfo{author}{\bibfnamefont{O.}~\bibnamefont{Skibbe}},
  \bibinfo{author}{\bibfnamefont{D.}~\bibnamefont{Vogel}},
  \bibinfo{author}{\bibfnamefont{M.}~\bibnamefont{Binder}},
  \bibinfo{author}{\bibfnamefont{A.}~\bibnamefont{Pucci}},
  \bibinfo{author}{\bibfnamefont{T.}~\bibnamefont{Kravchuk}},
  \bibinfo{author}{\bibfnamefont{L.}~\bibnamefont{Vattuone}},
  \bibinfo{author}{\bibfnamefont{V.}~\bibnamefont{Venugopal}},
  \bibinfo{author}{\bibfnamefont{A.}~\bibnamefont{Kokalj}}, \bibnamefont{and}
  \bibinfo{author}{\bibfnamefont{M.}~\bibnamefont{Rocca}}, \bibinfo{journal}{J.
  Chem. Phys.} \textbf{\bibinfo{volume}{131}}, \bibinfo{pages}{024701}
  (\bibinfo{year}{2009}).

\bibitem[{\citenamefont{Wang et~al.}(2004)\citenamefont{Wang, Jiang, Morikawa,
  Nakamura, Cai, Pan, and Zhao}}]{wang_SS04}
\bibinfo{author}{\bibfnamefont{G.-C.} \bibnamefont{Wang}},
  \bibinfo{author}{\bibfnamefont{L.}~\bibnamefont{Jiang}},
  \bibinfo{author}{\bibfnamefont{Y.}~\bibnamefont{Morikawa}},
  \bibinfo{author}{\bibfnamefont{J.}~\bibnamefont{Nakamura}},
  \bibinfo{author}{\bibfnamefont{Z.-S.} \bibnamefont{Cai}},
  \bibinfo{author}{\bibfnamefont{Y.-M.} \bibnamefont{Pan}}, \bibnamefont{and}
  \bibinfo{author}{\bibfnamefont{X.-Z.} \bibnamefont{Zhao}},
  \bibinfo{journal}{Surf. Sci.} \textbf{\bibinfo{volume}{570}},
  \bibinfo{pages}{205} (\bibinfo{year}{2004}).

\end{thebibliography}
\end{document}